\documentclass{jfm}
\input{Preamble}

\usepackage[usenames,dvipsnames]{color}
\newcommand{\ccc}{\textcolor{black}}
\newcommand{\cccc}{\textcolor{black}}

\title[Receptivity and sensitivity analysis of diffusion flames coupled with acoustics]{Global modes, receptivity, and sensitivity analysis of diffusion flames coupled with duct acoustics\thanks{This is a pre-print version. Published in J.\ Fluid\ Mech., vol. 752, 2014. Cambridge University Press$^{\copyright}$. DOI: http://dx.doi.org/10.1017/jfm.2014.328}}

\author[L. Magri and M. P. Juniper]
{Luca Magri\thanks{Email address for correspondence: \href{mailto:lm547@cam.ac.uk}{lm547@cam.ac.uk}} and Matthew P. Juniper}

\affiliation{Department of Engineering, University of Cambridge,\\
Trumpington Street, Cambridge, CB2 1PZ, UK}

\pubyear{xxxx}
\volume{xxx}
\pagerange{xxx--xxx}
\date{?? and in revised form ??}
\begin{document}
\maketitle
\begin{abstract}
In this theoretical and numerical paper, 
we derive the adjoint equations for a
thermo-acoustic system consisting of 
an infinite-rate chemistry diffusion flame coupled with duct acoustics. 
We then calculate the thermo-acoustic system's
linear global modes 
(i.e. the frequency/growth rate of oscillations, together with their mode shapes),
and the global modes'
receptivity to species injection,
sensitivity to base-state perturbations,
and structural sensitivity to advective-velocity perturbations.
Some of these could be found by finite difference calculations
but the adjoint analysis is computationally much cheaper.
We then compare these with the Rayleigh index.
The receptivity analysis shows 
the regions of the flame where open-loop injection of fuel or oxidizer 
will have most influence on the thermo-acoustic oscillation.
We find that the flame is most receptive at its tip. 
The base-state sensitivity analysis shows
the influence of each parameter on the frequency/growth rate. 
We find that perturbations to 
the stoichiometric mixture fraction, 
the fuel slot width,
and the heat-release parameter 
have most influence,
while perturbations to
the P\'eclet number have least influence. 
These sensitivities oscillate:
e.g. positive perturbations to the fuel slot width 
either stabilizes or destabilizes the system, depending on the operating point.
This analysis reveals that, as expected from a simple model,
the phase delay between velocity and heat-release fluctuations
is the key parameter in determining the sensitivities. 
It also reveals that this thermo-acoustic system
is exceedingly sensitive to changes in the base state.
The structural-sensitivity analysis shows
the influence of perturbations to the advective flame velocity.
The regions of highest sensitivity are around the stoichiometric line close to the inlet,
showing where velocity models need to be most accurate. 
This analysis can be extended to more accurate models
and is a promising new tool for the analysis and control of thermo-acoustic oscillations. 
\end{abstract}

\section{Introduction}
\par
In a thermo-acoustic system, 
heat-release oscillations couple with acoustic pressure oscillations in a feedback loop. 
If the heat released by the flame is sufficiently in phase with the pressure, the acoustic oscillations can grow \citep{Rayleigh1878}, sometimes with detrimental consequences to the performance of the system. 
These oscillations are a persistent problem. 
Their comprehension, prediction and control in the design of gas turbines and rocket engines are areas of current research, as reviewed by 
\citet{Lieuwen2005,Culick2006}.

\par
This theoretical and numerical paper examines the linear stability of a thermo-acoustic system. 
This system consists of an infinite-rate chemistry diffusion flame coupled with one-dimensional duct acoustics. 
The flame is assumed to be compact, meaning that it excites the acoustics as a pointwise heat source.
The heat-release is given by integration of the non-dimensional sensible enthalpy of the flame, which is solved in an ad-hoc two-dimensional domain.
This simple combustor was originally modelled by \citet{Tyagi2007b,Tyagi2007} using a finite-difference grid. 
We use a Galerkin method for discretization of the flame, however,
similar to that of \citet{Balasubramanian2008}. 
We reformulate the problem with revised equations \citep{Magri2013d} using a suitably normalized mixture fraction \citep{Peters1992,Poinsot2005}, so that the flame-acoustic coupled problem is well scaled, as suggested by \citet{Illingworth2013}. 
Also, we simulate the temperature discontinuity (or {\it jump}) in the mean flow,
which is caused by the heat released by the steady flame. 
This temperature jump induces
a discontinuous change in the speed of sound, 
which affects the thermo-acoustic modes' frequencies and wavelengths. 
We model this jump with a Galerkin method, drawing on the numerical model of \citet{Zhao2012}.

\par
The adjoint-based framework that we apply 
stems from ideas developed for 
the analysis of hydrodynamic instability
\citep{Hill1992,Chomaz2005,Giannetti2007}. 
\citet{Hill1992} and \citet{Giannetti2007} examined the flow behind a cylinder at $Re \approx 50$ and used this adjoint-based framework to reveal the region of the flow that causes von K\'{a}rm\'{a}n vortex shedding.
\citet{Giannetti2007} also used adjoint methods to calculate the effect that a small control cylinder has on the growth rate of oscillations, 
as a function of the control cylinder's position downstream of the main cylinder,
and compared this with experimental results by \citet{Strykowski1990}.
This analysis was further developed by \citet{Marquet2008} and \citet{LuchiniAIAA}, who considered the cylinder's effect on the base flow as well, which improved the comparison with experiments. 
Adjoint-based techniques have been applied to a large range of fluid dynamic systems,
most of which have been reviewed by \citet{Sipp2010} and \citet{Luchini2014}. 
Although \citet{Chandler2011} extended this analysis to low Mach number flows for variable density fluids and flames, adjoint equations have been used only recently in thermo-acoustics. 
\citet{Juniper2011} used nonlinear adjoint looping to find 
the nonlinear optimal states for triggering in a hot-wire Rijke tube. 
More recently, \citet{Magri2013e,MagriIJSCD,Magri2013} applied 
adjoint-based sensitivity analyses to this hot-wire Rijke tube. 
This paper extends these techniques
to the infinite-rate chemistry diffusion flame 
coupled with one-dimensional duct acoustics
in order to reveal
the most effective ways to change the 
stability/instability of the system.

\par
We describe the model in \S \ref{tamodel} 
and the numerical discretization in \S \ref{sec:discrete}.
The definition and derivation of the adjoint operator
and the general definition of the sensitivity are in \S \ref{adjoints}. 
In \S \ref{sec_res} we describe the most unstable mode of oscillation 
and interpret its driving mechanism with the Rayleigh Index. 
We then define and calculate 
(i) the system's receptivity to open-loop species injection in \S \ref{sec:rec_inj};
(ii) the system's sensitivity to changes in the combustion parameters in \S \ref{rad}, which are the stoichiometric mixture fraction, $Z_{sto}$, the fuel slot to duct width ratio, $\alpha$; the P\'eclet number, $Pe$; and the heat-release parameter, $\beta_T$;
(iii) the system's structural sensitivity to a generic advection-feedback mechanism in \S\ref{sec:rec_adv}.
These results are summarized in the conclusions. 
Further details about the methods are summarized in appendices \ref{appscale},\ref{appMeanFlow},\ref{ssbs},\ref{app:convergence} and in the online supplementary material.
\section{Thermo-acoustic model} \label{tamodel}
\par
\begin{figure}
\begin{center}
\includegraphics[width=0.9\textwidth, draft = false]{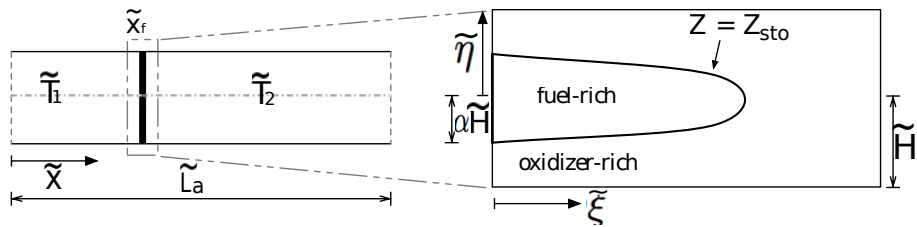}
\caption{Schematic of the dimensional thermo-acoustic model (dimensional quantities are denoted with $\tilde{}$ ). In non-dimensional variables (appendix \ref{appscale}) the acoustic domain is $[0, 1]$ and the flame domain $[0, L_c]\times[-1, 1]$.  This model is based on the compact-flame assumption, which means that the acoustic and flame  space domains are decoupled.} 
\label{model}
\end{center}
\end{figure}

The thermo-acoustic model consists of a diffusion flame placed in an  acoustic duct (figure~\ref{model}).
The acoustic waves
cause perturbations in
the velocity field.
In turn, these cause perturbations to the mixture fraction,
which convect down the flame
and cause perturbations
in the heat-release rate
and the dilatation rate at the flame. 
The dilatation described above provides a monopole source of sound,
which feeds into the acoustic energy. 
We assume that the flame is compact,
meaning that the heat release is a point-wise impulsive forcing term for the acoustics. 
\cccc{
One limitation of this model is that the velocity in the flame domain is assumed to be uniform in space. This allows for convection, as described above, but does not allow for flame wrinkling or pinch-off. Another limitation is that the infinite rate chemistry does not permit flame blow-off, which is another source of heat release oscillations at large oscillation amplitudes. Neither limitation will have much influence on a linear study such as this, however, because perturbations are infinitesimal and therefore wrinkling, pinch-off, and blow-off will not occur. They would, however, be important for a nonlinear study.
} %

\subsection{Acoustic model}\label{acmodel}
\par
We model one-dimensional
acoustic velocity and pressure perturbations, $u$ and $p$,
on top of an inviscid flow with Mach number $\lesssim0.1$. 
Under these assumptions, we can neglect the effect of the mean-flow velocity (see figure \ref{fig:compWG} in appendix \ref{appMeanFlow}).
The flame causes discontinuities in the mean-flow density and speed of sound, 
which we model with a Galerkin method.
The acoustics are governed by the momentum and energy equations, respectively
\begin{eqnarray}
\label{equ_gov_mom_ndim}
&&\rho \frac{\partial u}{\partial t} + \frac{\partial p}{\partial x} = 0 ,\\
\label{equ_gov_enr_ndim}
&&\frac{\partial p}{\partial t} + \frac{\partial u}{\partial x}
+ \zeta p
- \beta_T\dot{Q}_{av}\delta(x - x_f) = 0, 
\end{eqnarray}
where $\rho$, $u$, $p$ and $\dot{Q}_{av}$ are the non-dimensional density, velocity, pressure, and heat-release rate integrated over the combustion domain.
We label these the \emph{direct} equations.
The characteristic scales used for non-dimensionalization are in appendix \ref{appscale}.
The acoustic base-state parameters, which we can control, are 
$\zeta$, which is the damping; 
$x_f$, which is the flame position; and the heat-release parameter, 
$\beta_T=1/T_{av}=2/(T_1+T_{ad})$, 
where
$T_1$ is the reactants' inlet temperature 
and $T_{ad}$ is the adiabatic flame temperature.
With the mixture fraction formulation adopted in this paper \citep{Poinsot2005}, $T_{ad}=Z_{sto}/2$, where $Z_{sto}$ is the stoichiometric mixture fraction defined afterwards in \S\ref{flame_model}, eq. (\ref{Zsto}). 
The system \eqref{equ_gov_mom_ndim}, \eqref{equ_gov_enr_ndim} reduces to the D'Alembert equation when $\zeta=0$ and $\beta_T=0$
\begin{equation}\label{dalembert}
\frac{\partial^2 p}{\partial t^2} - \frac{1}{\rho}\frac{\partial^2 p}{\partial x^2}=0.
\end{equation}
\par
The non-dimensional mean-flow density, $\rho$, 
is modelled as a discontinuous function
\begin{eqnarray}\label{rho_heavi}
\rho= 
\begin{cases}
\rho_1, & 0\leq x<x_f,\\
\rho_2, &x_f<x\leq1.
\end{cases}
\end{eqnarray}
The densities can be obtained from the temperatures, 
which are $\tilde{T}_1$
in the cold flow upstream of the flame, 
$0\leq x<x_f$,
and $\tilde{T}_2$
in the hot flow downstream of the flame, 
$x_f<x\leq1$, by the first law of thermodynamics and ideal-gas state equation:
\begin{equation}\label{thermo_law}
\frac{\rho_1}{\rho_2} 
= \frac{T_2}{T_1}
= \frac{\tilde{T_2}}{\tilde{T_1}}
= 1 + \frac{\tilde{\bar{Q}}}{\tilde{c_p} \tilde{T_1}},
\end{equation}
where  $\tilde{\bar{Q}}$ is the steady heat release; 
$\tilde{c_p}$ is the constant-pressure heat capacity;
and $\tilde{}$ indicates a dimensional quantity. 
This has assumed that the mean-flow pressure drop across the flame is negligible,
which is reasonable when $\gamma M_1^2$ and $\gamma M_2^2$ are small \citep[see e.g.][]{Dowling1997}, where $\gamma$ is the heat-capacity ratio and $M_1,M_2$ are the mean-flow Mach numbers. 
\par
At the ends of the tube, $p$ and $\partial u/ \partial x$ are both set to zero, which means that the system cannot dissipate acoustic energy by doing work on the surroundings. Dissipation and end losses are modelled by the modal damping  $\zeta=c_1j^2+c_2j^{0.5}$ used by \citet{Matveev2003a}, based on models by \citet{Landau1987}, where $j$ is the $j^{th}$ acoustic mode \cccc{and $c_1$, $c_2$ are the constant damping coefficients.} The quadratic term represents the losses at the end of the tube, while the square-rooted term represents the losses in the viscous/thermal boundary layer. 
\subsection{Flame model} \label{flame_model}
\par
In the flame domain (right picture of figure \ref{model}),
the fuel enters the left boundary at $-\alpha\leq\eta\leq\alpha$
and the oxidizer enters the left boundary at 
$-1\leq\eta\leq-\alpha$ and $\alpha\leq\eta\leq\ 1$.
The main assumptions are that:
(i) the velocity and density in the flame domain are uniform;
(ii) the Lewis number, defined as the ratio of thermal diffusivity to mass diffusivity, is 1;
(iii) the mass-diffusion coefficients are isotropic and uniform;
(iv) the chemistry is infinitely fast with one-step reaction. 
We define the mass fraction to be 
the mass of a species divided by the total mass of the mixture (kg/kg).
The fuel mass fraction is labelled $Y^*$
and the oxidizer mass fraction is $X^*$.
The stoichiometric mass ratio is 
$s = \nu_X W_X / (\nu_Y W_Y)$,
where $W_X$ and $W_Y$ are the molar masses (kg/mole) 
and $\nu_X$ and $\nu_Y$ are the stoichiometric coefficients (mole/kg).
We define a conservative scalar, \emph{Z}, called the mixture fraction \citep{Peters1992,Poinsot2005}
\begin{align} \label{Zvar}
Z \equiv \frac{sY^* - X^* +X^*_i}{sY^*_i + X^*_i}=\frac{Y-X+X_i}{X_i+Y_i},
\end{align}
where $\mathit{X} = X^*/(\nu_XW_X)$ and $\mathit{Y} = Y^*/(\nu_YW_Y)$,
and the subscript \emph{i} refers to properties evaluated at the inlet. 

Earlier definitions of the mixture fraction \citep{Tyagi2007,Balasubramanian2008,Magri2013d}, depended on the absolute value of the fuel mass fraction, $Y_i$. This dependency has been overcome by defining $Z$ as in \eqref{Zvar}, which can assume only values between 0 and 1, 
rendering the non-dimensionalization of the coupled themo-acoustic system well scaled. 
This flame formulation has been used to characterize the nonlinear thermo-acoustic behaviour of ducted diffusion flames by \citet{Illingworth2013}.
\par
The fuel and oxidizer  diffuse into each other 
and, under the infinite-rate chemistry assumption, combustion occurs in an infinitely thin region at the stoichiometric contour, $Z = Z_{sto}$, where 
\begin{align} \label{Zsto}
Z_{sto} = \frac{1}{1+\phi},
\end{align}
where $\phi \equiv Y_i/X_i$ is the \emph{equivalence ratio} \citep[eq.~(3.17), p. 86]{Poinsot2005}. 
The governing equation for $Z$ is derived from the species equations
\citep{Tyagi2007,Tyagi2007b,Balasubramanian2008}
and, in non-dimensional form, is 
\begin{align} \label{eq:Z}
&\frac{\d Z}{\d t} + (1+u_f)\frac{\d Z}{\d \xi} - \frac{1}{\Pen}\left(\frac{\d^2 Z}{\d \xi^2} + \frac{\d^2 Z}{\d \eta^2}\right)=0,
\end{align}
\cccc{where $1$ is the non-dimensional mean-flow velocity (see appendix \ref{appscale} for the scale factors used)}; 
$u_f$ is the acoustic velocity evaluated at the flame location; and
$\Pen$ is the P\'eclet number (defined in appendix \ref{appscale}).  
The partial differential equation \eqref{eq:Z} is  parabolic and, when the flame is coupled with acoustics, quasilinear. 
Dirchlet boundary conditions are prescribed at the inlet
\begin{align}
&Z(\xi=0,\eta)= 1 \qquad\textrm{if $|\eta | \leq\alpha$}, \label{bc1}\\
&Z(\xi=0,\eta)= 0 \qquad\textrm{if $\alpha<|\eta | \leq 1$}. \label{bc2} 
\end{align}
These assume that axial back diffusion at $\xi=0$ is negligible, which is a good assumption for the P\'eclet numbers we investigate \citep{Magina2014}.
Neumann boundary conditions are prescribed elsewhere
\begin{align}
&\frac{\partial Z}{\partial \eta}(\xi,\eta=\pm1)=0, \label{bc3} \\ \label{bc4}
&\frac{\partial Z}{\partial \xi}(\xi=L_c,\eta)=0. 
\end{align}
These ensure that there is no diffusion across the upper and lower wall of the combustor, and that $Z$ is uniform at the end of the flame domain. 
\par
The variable $Z$ is split into two components, $Z=\bar{Z}+ z$,
in which $\bar{Z}$ is the steady solution,
\begin{equation}\label{Zbar}
\bar{Z} = \frac{\bar{Y}-\bar{X}+X_i}{X_i+Y_i} = \frac{\bar{Y}-\bar{X}}{X_i+Y_i} + Z_{sto},
\end{equation}
and $z$ is the unsteady field, 
\begin{equation}\label{zunste}
z = \frac{y-x}{X_i+Y_i}.
\end{equation}
By decomposition \eqref{Zbar} and \eqref{zunste}, the mixture-fraction equation \eqref{eq:Z} is split into a steady and fluctuating part governed by
\begin{align} \label{eq:Zst}
&\frac{\partial \bar{Z}}{\partial \xi}-\frac{1}{\Pen}\left(\frac{\d^2 \bar{Z}}{\d \xi^2} + \frac{\d^2 \bar{Z}}{\d \eta^2}\right)=0, \\
&\frac{\partial z}{\partial t}-\frac{1}{\Pen}\left(\frac{\d^2 z}{\d \xi^2} + \frac{\d^2 z}{\d \eta^2}\right)+\left(1+u_f\right)\frac{\partial z}{\partial \xi}=-u_f\frac{\partial \bar{Z}}{\partial \xi}.\label{eq:zql}
\end{align}
The steady field, $\bar{Z}$, has the same boundary condition as the variable \emph{Z}, given in (\ref{bc1})--(\ref{bc4}). Equation (\ref{eq:Zst}) has an analytical solution \citep{Magri2013e,Magina2013}, which is reported in appendix~\ref{ssbs}.
The unsteady component, \emph{z}, must satisfy the Neumann boundary conditions (\ref{bc3}), (\ref{bc4}) but must be zero at the inlet, $\xi=0$.
In order to linearize \eqref{eq:zql}, we assume that $z\sim u_f\sim O(\epsilon)$, so that we 
discard the term $u_f\partial z/\partial\xi\sim O(\epsilon^2)$, yielding
\begin{align} 
\frac{\partial z}{\partial t}-\frac{1}{\Pen}\left(\frac{\d^2 z}{\d \xi^2} + \frac{\d^2 z}{\d \eta^2}\right)+\frac{\partial z}{\partial \xi}=-u_f\frac{\partial \bar{Z}}{\partial \xi}.\label{eq:z1}
\end{align}
\subsection{Heat-release model}\label{sec:heat}
The non-dimensional heat release (rate) is given by the integral
of the total derivative of the non-dimensionalized sensible enthalpy
\begin{align} \label{Qd}
&\dot{Q} = \int_{R}\frac{\mathrm{d}(T_b-T_i)}{\mathrm{d}t}\mathrm{d}\xi\mathrm{d}\eta,\\
&T_b = T_i + \bar{Z} + z, &\textrm{if Z $<$ $Z_{sto}$},\label{To}\\ \label{Tf}
&T_b = T_i + \frac{Z_{sto}}{1-Z_{sto}}\left(1- \bar{Z} -z \right), &\textrm{if Z$\ge$ $Z_{sto}$},
\end{align}
where $R\equiv[0, L_c]\times[-1, 1]$ is the flame domain, and
$T_i$ is the non-dimensional inlet temperature of both species.
Note that, following the notation used for the acoustics in \S\ref{acmodel}, $T_i\equiv T_1$. 
The value of the steady heat release rate, $\bar{Q}$, 
depends on whether the flame is closed (overventilated), $Z_{sto}>\alpha$,
or open (underventilated), $Z_{sto}<\alpha$
\begin{align}
&\bar{Q} = 2\alpha - \frac{1}{1-Z_{sto}}\int_{-1}^{+1}z(L_c,\eta)\mathrm{d}\eta&\textrm{if $Z_{sto}\geq\alpha$}, \label{Qstclose}\\
&\bar{Q} = 2\left(\frac{Z_{sto}}{1-Z_{sto}}\right)\left(1-\alpha\right) &\textrm{if $Z_{sto}\leq\alpha$}.\label{Qstopen}
\end{align}
The fluctuating heat-release, integrated over the flame domain, is 
\begin{align}\label{eq:Qav}
\dot{q}_{av}&\equiv\dot{Q}-\bar{Q} = \nonumber\\
&=\int_0^{L_c}\int_{-1}^1
\left\{
\mathrm{\Theta}(Z\geq Z_{sto})\left(\frac{-Z_{sto}}{1-Z_{sto}}\right)\frac{\partial z}{\partial t} +
\mathrm{\Theta}(Z< Z_{sto})\frac{\partial z}{\partial t}
\right\}
\;\mathrm{d}\xi\mathrm{d}\eta
+ u_f\bar{Q},
\end{align}
where $\mathrm{\Theta}(Z\geq Z_{sto})$ is 1 in the fuel side ($Z\geq Z_{sto}$) and zero otherwise,
and
$\mathrm{\Theta}(Z< Z_{sto})$ is 1 in the oxidizer side ($Z< Z_{sto}$) and zero otherwise. 
Numerical calculations show that the term $\int_{-1}^{+1}z(L_c,\eta)\mathrm{d}\eta$ in \eqref{Qstclose} is negligible, being of order $\sim 10^{-13}$.
Expression \eqref{eq:Qav} is valid for both closed and open flames. 
The heat release (\ref{eq:Qav}) has to be scaled further in order to be consistent with the non-dimensionlization of the acoustic energy (\ref{equ_gov_enr_ndim}). 
Bearing in mind that the dimensional width of the duct is $2\tilde{H}$ (figure~\ref{model}) and considering the scale factors in appendix~\ref{appscale}, then the heat-release term
forcing the acoustic energy (\ref{equ_gov_enr_ndim}) is $\dot{Q}_{av}=\dot{q}_{av}/2$.
\section{Numerical discretization}\label{sec:discrete}
\par
Both the acoustics and flame are discretized with the Galerkin method. 
The partial differential equations \eqref{equ_gov_mom_ndim}, \eqref{equ_gov_enr_ndim}, \eqref{eq:zql}
are discretized into a set of ordinary differential equations
by choosing a basis that matches the boundary conditions and the discontinuity condition at the flame. 
The Galerkin method, which is a weak-form method, 
ensures that the error is orthogonal to the chosen basis
in the subspace in which the solution is discretized, 
so that the solution is an optimal weak-form solution.
The pressure, $p$, and velocity, $u$, are expressed by separating the time and space dependence, as follows
\begin{eqnarray}\label{equ_gal_p1}
p(x,t)=\sum_{j = 1}^{K} 
\begin{cases}
\alpha_j^{(1)}(t)\Psi^{(1)}_j(x), & 0\leq x<x_f,\\
\alpha_j^{(2)}(t)\Psi^{(2)}_j(x), &x_f<x\leq1,
\end{cases}
\end{eqnarray}
\begin{eqnarray}\label{equ_gal_u1}
u(x,t)=\sum_{j = 1}^{K} 
\begin{cases}
\eta_j^{(1)}(t)\Phi^{(1)}_j(x), & 0\leq x<x_f,\\
\eta_j^{(2)}(t)\Phi^{(2)}_j(x), &x_f<x\leq1.
\end{cases}
\end{eqnarray}
The following procedure is applied to find the bases for $u$ and $p$:
\begin{enumerate} 
\item substitute the decomposition \eqref{equ_gal_p1} into \eqref{dalembert} to find 
the acoustic pressure eigenfunctions $\Psi^{(1)}_j(x)$, $\Psi^{(2)}_j(x)$;
\item  substitute the pressure eigenfunctions  $\Psi^{(1)}_j(x)$, $\Psi^{(2)}_j(x)$ into the momentum equation \eqref{equ_gov_mom_ndim} to find the acoustic velocity eigenfunctions  $\Phi^{(1)}_j(x)$, $\Phi^{(2)}_j(x)$;
\item impose the jump condition at the discontinuity, for which $p(x\rightarrow x^-_f)=p(x\rightarrow x^+_f)$ and $u(x\rightarrow x^-_f)=u(x\rightarrow x^+_f)$ \citep[see e.g.][]{Dowling2003}, to find the relations between $\alpha_j^{(1)}$, $\alpha_j^{(2)}$, $\eta_j^{(1)}$, $\eta_j^{(2)}$. 
\end{enumerate}
 Similarly to  \citet{Zhao2012}, these steps give
 \begin{eqnarray}\label{equ_gal_p11}
p(x,t)=\sum_{j = 1}^{K} 
\begin{cases}
-\alpha_j(t)\sin\left(\omega_j\sqrt{\rho_1}x\right), & 0\leq x<x_f,\\
-\alpha_j(t)\left(\frac{\sin\gamma_j}{\sin\beta_j}\right)\sin\left(\omega_j\sqrt{\rho_2}(1-x)\right), &x_f<x\leq1,
\end{cases}
\end{eqnarray}
\begin{eqnarray}\label{equ_gal_u11}
u(x,t)=\sum_{j = 1}^{K} 
\begin{cases}
\eta_j(t)\frac{1}{\sqrt{\rho_1}}\cos\left(\omega_j\sqrt{\rho_1}x\right), & 0\leq x<x_f,\\
- \eta_j(t)\frac{1}{\sqrt{\rho_2}}\left(\frac{\sin\gamma_j}{\sin\beta_j}\right)\cos\left(\omega_j\sqrt{\rho_2}(1-x)\right), &x_f<x\leq1.
\end{cases}
\end{eqnarray}
where
\begin{equation}\label{cosgcosb}
\gamma_j \equiv \omega_j\sqrt{\rho_1}x_f, \;\;\;\;\beta_j \equiv\omega_j\sqrt{\rho_2}(1-x_f).
\end{equation}
Point {\it(c)} of the previous procedure provides the equation for the acoustic angular frequencies $\omega_j$
\begin{equation}\label{eig_val}
\sin\beta_j\cos\gamma_j+\cos\beta_j\sin\gamma_j\sqrt{\frac{\rho_1}{\rho_2}}=0.
\end{equation}
\begin{figure}
\begin{center}
\includegraphics[width=0.6\textwidth, draft = false]{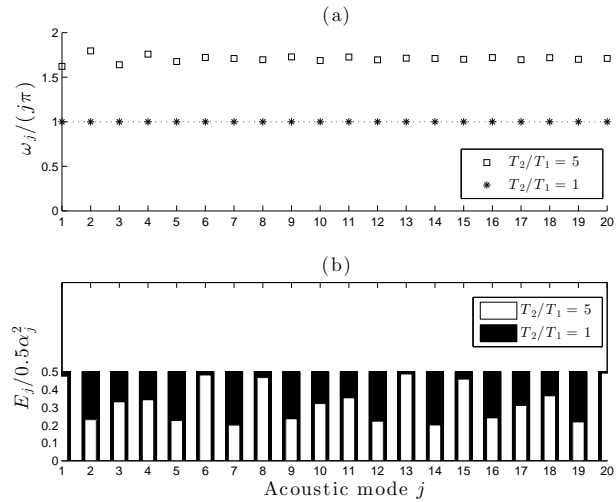}
\caption{The effect of the temperature jump on (a) the angular frequency  and (b) the pressure energy of  each mode of the undamped acoustic system (with no unsteady heat source). The flame position is $x_f=0.25$. The presence of the temperature jump markedly affects the acoustic frequencies and the modal distribution of the pressure energy.} 
\label{Jump}
\end{center}
\end{figure}
\begin{figure}
\begin{center}
\includegraphics[width=0.9\textwidth, draft = false]{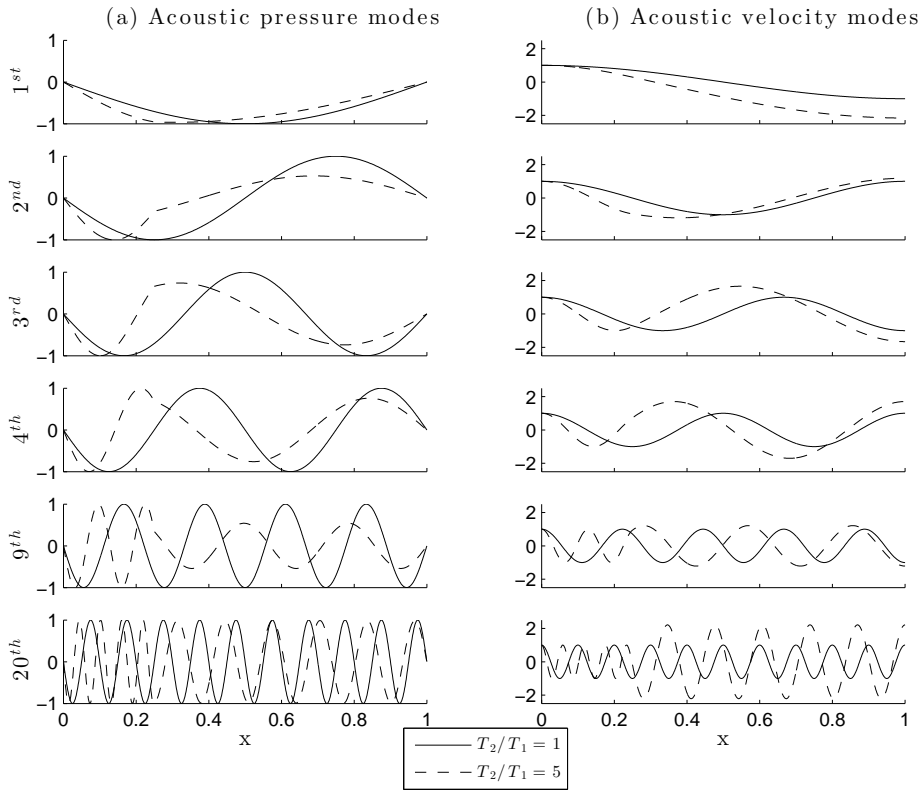}
\caption{The acoustic eigenfunctions of (a) pressure and (b) velocity with temperature jump (dashed lines) and without temperature jump (solid lines). The flame position is $x_f=0.25$.} 
\label{AcModes}
\end{center}
\end{figure}
Note that in the limit $\rho_1=\rho_2$, we recover the Galerkin expansion for a flow with no discontinuity of the mean properties across the flame \citep[e.g.][]{Balasubramanian2008a,Balasubramanian2008}. 
Importantly, in this limit the angular frequencies of the acoustic eigenfunctions are $\omega_j=j\upi$ (figure \ref{Jump}a). 
Such a limit is justified when the temperature jump is sufficiently low, i.e. $T_2/T_1\lesssim1.5$ \citep{Heckl1988,Dowling2005}. 
On the other hand, when the temperature jump is higher, as in realistic combustors, we have to consider the effect of the change of mean properties on the shape and frequency of oscillations \citep{Dowling1995}.
When the discontinuity is high, i.e. $T_2/T_1=5$ \citep{Nicoud2009}, the fundamental angular frequency is almost 1.6-1.8 times that of the case with no discontinuity, as depicted in figure \ref{Jump}a. 
The quantity $E_j$ in figure \ref{Jump}b, which originates from the projection of the energy equation \eqref{equ_gov_enr_ndim} along the Galerkin basis \eqref{equ_gal_p1}, is physically the acoustic-pressure energy stored in the $j$th mode, scaled by $0.5\alpha_j^2$. 
The system with no discontinuity has a constant acoustic pressure energy distribution with no dependence on the acoustic mode. When the discontinuity is modelled, however, the acoustic-pressure energy is mode-dependent and always lower (figure \ref{Jump}b) than it is in the system with no temperature jump. 
The acoustic modes, which are the basis functions for the Galerkin method, are markedly affected by the presence of the temperature jump (figure \ref{AcModes}). When the temperature jump is present, the acoustic wavelength rises across the discontinuity, as inferable from \eqref{equ_gal_p11},\eqref{equ_gal_u11}. In addition, the effect that the mean-flow velocity has on the acoustic angular frequencies is negligible for $M_1\lesssim0.1$, as reported in figure \ref{fig:compWG} in appendix \ref{appMeanFlow}.

By the Galerkin method, the flame is discretized as
\begin{equation}
z(\xi, \eta, t) = \sum_{m=1}^M\sum_{n=0}^{N-1} G_{n,m}(t)\cos(n\upi \eta)\sin\left[\left(m-\frac{1}{2}\right)\frac{\upi \xi}{L_c}\right].\label{zGalerk}%
\end{equation}
Via discretization \eqref{zGalerk}, the space resolution of the flame is half the shortest wavelength, which is $L_c/(M-0.5)$ in the $\xi$-direction and $1/N$ in the $\eta$-direction. (Although the flame mode $n=0$ is necessary to make the basis complete, in our calculations we noticed that this mode is negligible, having no effect on the system's dynamics and stability.)
\par
The state of the discretized system is defined by 
the amplitudes of the Galerkin modes that represent 
the flame, $\mathit{G_{n,m}}$, 
the velocity, $\mathit{\eta_j}$,
and the pressure, $\mathit{\alpha_j}$. 
These are collected in the column vector
$\boldsymbol{\chi} \equiv (\boldsymbol{G}; \boldsymbol{\eta}; \boldsymbol{\mathit{\alpha}}) $, where $\boldsymbol{G}\equiv (G_{0,1}; G_{0,2};\ldots; G_{1,1}; \ldots; G_{NM,NM})$; $\boldsymbol{\eta} \equiv  (\eta_1; \eta_2; \ldots; \eta_K)$; and $\boldsymbol{\alpha} \equiv  (\alpha_1; \alpha_2; \ldots; \alpha_K)$. 
Therefore the Galerkin-discretized thermo-acoustic system can be represented in state-space formulation as
\begin{equation}
\mathsfbi{M}\frac{\mathrm{d} \boldsymbol{\chi}}{\mathrm{d} t} = \mathsfbi{B} \boldsymbol{\chi} - u_f\mathsfbi{A}\boldsymbol{\chi}, \label{state_matrix}
\end{equation}
where $\mathsfbi{M}$, $\mathsfbi{B}$ and $\mathsfbi{A}$ are $\left(NM+2K\right)\times\left(NM+2K\right)$ matrices (\cccc{all of which are invertible}, see online supplementary material)
and $\boldsymbol{\chi}$ is the  $\left(NM+2K\right)\times1$ state vector. The term $u_f\mathsfbi{A}\boldsymbol{\chi}$ is the quasilinear term, which is discarded in linear analysis \eqref{eq:z1}. 

The strength of the Galerkin method is that the system can be expressed in state-space formulation \eqref{state_matrix}. This is particularly useful when the adjoint algorithm is to be implemented. Other numerical discretizations, such as Chebyshev polynomials used by \citet{Illingworth2013}, are numerically more efficient but make the implementation of the adjoint problem much more difficult because of the way that the integration of the heat release is handled. 
It is worth pointing out that \citet{Sayadi2013} have developed a new numerical method for the acoustics that, amongst other things, prevents the Gibbs' phenomenon across the discontinuity, which arises with a fine Galerkin discretization of the acoustics \citep{Magri2013}.
\section{Adjoint analysis} \label{adjoints}
\subsection{Adjoint operator}\label{adj_def_sec}
\par
In this section the \emph{adjoint operator} is defined. 
This definition is an extension for functions (arranged in vector-like notation) over the time domain of the definition given by \citet{Dennery1996}.
Let $\mathrm{L}$ be a partial differential operator of order $\mathit{M}$ acting on the function $\mathbf{q}(x_1,x_2,\ldots,x_K,t)$, where $\mathit{K}$ is the space dimension, such that $\mathrm{L}\mathbf{q}(x_1,x_2,\ldots,x_K,t)=0$. 
We refer to the operator $\mathrm{L}$ as the \emph{direct operator} and the function $\mathbf{q}$ as the \emph{direct variable}.
The adjoint operator $\mathrm{L}^+$ and adjoint variable $\mathbf{q}^+(x_1,x_2,\ldots,x_K,t)$ are defined via the \emph{generalized Green\rq{}s identity}:
\begin{align}
&\int_0^T \! \left\langle \mathbf{q}^{+}, \mathrm{L}\mathbf{q}\right\rangleÊ- \left\langle\mathbf{q}, \mathrm{L}^+\mathbf{q}^+\right\rangle \mathrm{d}t=\nonumber \\
&=\int_0^T \int_{S}\!\; \sum_{i=1}^{K} \left[ \frac{\partial}{\partial x_i} F_i \left(\mathbf{q},\mathbf{q}^{+*}\right) \right] n_i \mathrm{d}S \mathrm{d}t + \int_{V}Q\left(\mathbf{q},\mathbf{q}^{+*}\right)\rvert_0^T\mathrm{d}V,\label{Greens_gen}
\end{align}
where $\mathit{i}=1,2,\ldots,K$. $F_i(\mathbf{q},\mathbf{q}^{+*})$, which is referred as the \emph{bilinear concomitant} \citep[see e.g.][]{Giannetti2007}, and $Q\left(\mathbf{q},\mathbf{q}^{+*}\right)$, which is a functional, depend bilinearly on $\mathbf{q}$, $\mathbf{q^{+*}}$ and their first $\mathit{M-1}$ derivatives. 
The complex-conjugate operation is labelled by $^*$. 
For brevity, we define $\langle\mathbf{a},\mathbf{b}\rangle\equiv\int_V \mathbf{a}^*\mathbf{\cdot}\mathbf{b}\;\mathrm{d}V$, where $\mathbf{a}$, $\mathbf{b}$ are suitably differentiable vector functions; and the Euclidean scalar product is indicated with the dot $\;\;\mathbf{\cdot}\;\;$. (We choose to define the adjoint equation through an inner product, but any non-degenerate bilinear form could have been used.)
The domain $\mathit{V}$ is enclosed by the surface $S$, for which $\mathit{n_i}$ are the projections onto the coordinate axis of the unit vector in the direction of the outward normal to the surface $\mathrm{d}S$. 
The time interval is $\mathit{T}$. 
The adjoint boundary and initial conditions on the function $\mathbf{q}^+$ are defined as those that make the right-hand side of (\ref{Greens_gen}) vanish identically on $S$, $\mathit{t}=0$ and $\mathit{t=T}$.

The adjoint equations can either be derived from the continuous direct equations and then discretized ($\CA$, discretization of the Continuous Adjoint) or be derived directly from the discretized direct equations ($\DA$, Discrete Adjoint).
%
For the $\CA$ method, 
the adjoint equations are derived by
integrating the continuous direct equations by parts
and then applying the Green\rq{}s identity (\ref{Greens_gen}).
For the $\DA$ method
the adjoint system is the negative Hermitian of the direct system. 
This can be obtained algorithmically by reverse routine-calling \citep{Errico1997,Bewley2001,Luchini2014}. 
Generally, the $\DA$ method has the same truncation errors
as the discretized direct system,
while the $\CA$ method has different truncation errors, depending on the choice of the numerical discretization \citep{Vogel1995,Magri2013}. 
The continuous adjoint equations of the linear thermo-acoustic system, consisting of (\ref{equ_gov_mom_ndim}), (\ref{equ_gov_enr_ndim}), 
are:
\begin{equation} \label{adl13}
\frac{\partial u^+_f}{\partial t} + \frac{\partial p^+_f}{\partial x} - z^+\frac{\partial \bar{Z}}{\partial \xi} +\left(\frac{1}{1-Z_{sto}}\right) \dot{q}^+ \bar{Q}   =0 ,
\end{equation}
\begin{equation} \label{adl23}
\frac{\partial u^+}{\partial x} + \frac{\partial p^+}{\partial t} - \zeta p^+=0 ,
\end{equation}
\begin{equation} \label{adl33}
p^+\delta(x-x_f) - {\beta_T}\dot{q}^+=0 ,
\end{equation}
\begin{equation} \label{adl43}
\frac{\partial z^+}{\partial t}+ \bar{U}\frac{\partial z^+}{\partial \xi} -\frac{1}{\Pen}\left(\frac{\partial^2z^+}{\partial \xi^2}+\frac{\partial^2z^+}{\partial \eta^2}\right) +\left(\frac{1}{1-Z_{sto}}\right) A(\bar{Z}) \frac{\partial \dot{q}^+}{\partial t} =0,
\end{equation}
where $u^+=u^+(x,t)$, $p^+=p^+(x,t)$ and $z^+=z^+(\xi,\eta,t)$.
The area enclosed by the steady stoichiometric line is labelled $A(\bar{Z})$. 
The adjoint boundary conditions are
\begin{align}
p^+=0 \qquad&\textrm{at\;\;}x=0,\;\;x=1,\label{pbc}\\
z^+=0 \qquad&\textrm{at\;\;}\xi=0, \label{tc1}\\
\nabla z^+\mathbf{\cdot}\mathbf{n}=0\qquad&\textrm{at\;\;} \xi=L_c, \;\eta=\pm1. \label{tc2}
\end{align}
\par
\cccc{The adjoint boundary conditions \eqref{tc1}-\eqref{tc2} are the same as those of the direct problem, which means that the basis used in \eqref{zGalerk} is suitable for spanning the flame's adjoint space.}

The adjoint equations govern the evolution of the adjoint variables, which can be regarded as Lagrange multipliers from a constrained optimization perspective \citep{Belegundu1985,Gunzbur,Giles2000}. 
Therefore, $u^+$ is the Lagrange multiplier of the acoustic momentum equation (\ref{equ_gov_mom_ndim}), revealing the spatial distribution of the acoustic system's sensitivity to a force. Likewise, $p^+$ is the Lagrange multiplier of the pressure equation (\ref{equ_gov_enr_ndim}), revealing the spatial distribution of the acoustic system's sensitivity to heat injection.
Finally, $z^+$ is the Lagrange multiplier of the flame equation (\ref{eq:zql}), revealing the spatial distribution of the combustion system's sensitivity to species injection (\S\ref{sec:rec_inj}). 
A mathematical treatment of the adjoint equations, interpreted for thermo-acoustics, is given by \citet{MagriIJSCD}.
\par
For linear thermo-acoustic systems arranged in a state-space formulation, such as system (\ref{state_matrix}), the DA method is more accurate and easier to implement than the CA method (see, for example, \citet{Magri2013}).
Therefore, we will use the DA method in this paper.
\par
So far 
we have considered the thermo-acoustic system in the ($\mathit{x}$, $\xi$, $\eta$, $\mathit{t}$) domain.
In modal analysis, 
we consider it in the ($\mathit{x}$, $\xi$, $\eta$, $\mathit{\sigma}$) domain using the modal transformations
$u(x,t)=\hat{u}(x,\sigma)\exp(\sigma t)$,  $p(x,t)= \hat{p}(x,\sigma)\exp(\sigma t)$, and
$z(\xi,\eta,t)=\hat{z}(\xi,\eta,\sigma)\exp(\sigma t)$.  
The symbol $\;\hat{ }\;$ denotes an eigenfunction. 
The complex eigenvalue is $\sigma = \sigma_r +  \sigma_i\textrm{i}$, where $(\sigma_r,\sigma_i) \in \mathbb{R}^2$.
The behaviour of the system in the long-time limit 
is dominated by the eigenfunction 
whose eigenvalue has the highest real part (i.e. growth rate), $\sigma_r$.\\
\subsection{Sensitivity}\label{sec:str_th}
Adjoint eigenfunctions are useful because they provide gradient information about the sensitivity of the system's stability to first-order perturbations to the governing operator.
Defining the operator in \S \ref{adj_def_sec} as $\mathrm{L}\equiv\mathrm{M}\partial/\partial t - \mathrm{B}$, the continuous generalized eigenproblem of \eqref{state_matrix} \cccc{and its adjoint} are, respectively 
\begin{align} \label{genop}
&\sigma\mathrm{M}\hat{\mathbf{q}}=\mathrm{B}\hat{\mathbf{q}},\\ \label{genopa}
&\cccc{\sigma^*\mathrm{M}^+\hat{\mathbf{q}}^+=\mathrm{B}^+\hat{\mathbf{q}}^+},
\end{align}
where $\mathrm{M}$ may be a non-invertible matrix of operators. 
\cccc{The adjoint operators $\mathrm{M}^+$ and $\mathrm{B}^+$ can be regarded as the conjugate transpose of the corresponding direct operators, $\mathrm{M}$ and $\mathrm{B}$, respectively.}
The sensitivity of the eigenvalues
to generic perturbations to the system
can be obtained by introducing a perturbation operator, $\delta\textrm{C}\cdot\mathrm{P}$,
such that the perturbed operator is 
$\mathrm{B}\rightarrow\mathrm{B}+\delta \textrm{C}\cdot\mathrm{P}$, where $\delta\textrm{C}$ is a gain operator and $\textrm{P}$ is the perturbation operator. The gain is small such that its (suitably defined) norm is $\lvert\lvert\delta\mathrm{C}\lvert\lvert=\lvert\epsilon\lvert\ll1$. 
This perturbation changes the eigenvalues and eigenfunctions accordingly:
$\sigma\rightarrow\sigma+\epsilon\delta\sigma$, 
$\hat{\mathbf{q}} \rightarrow \hat{\mathbf{q}} +\epsilon\delta \hat{\mathbf{q}}$,
and 
$ \hat{\mathbf{q}}^+ \rightarrow \hat{\mathbf{q}}^+ +\epsilon\delta \hat{\mathbf{q}}^+$. 
By retaining only first-order terms $\sim O(\epsilon^1)$,
and taking into account the bi-orthogonality condition \citep{Salwen1981}, 
the sensitivity of the eigenvalue 
is calculated as follows 
\begin{equation}\label{str_sens_s}
\frac{\delta\sigma}{\delta\textrm{C}} = \frac{\left\langle \hat{\mathbf{q}}^+,  \textrm{P}\hat{\mathbf{q}}\right\rangle}{\left\langle \hat{\mathbf{q}}^+, \mathrm{M}\hat{\mathbf{q}}\right\rangle}.
\end{equation}
This result is well known from spectral theory and was used for the first time in flow instability by \citet{Hill1992} and \citet{Giannetti2007}. 
For the thermo-acoustic system in this study,
the eigenfunctions are arranged in column vectors as $\hat{\mathbf{q}}\equiv\left[\hat{z}; \hat{u}; \hat{p}\right]$, $\hat{\mathbf{q}}^+\equiv\left[\hat{z}^{+}; \hat{u}^{+}; \hat{p}^{+}\right]$; the integration domain is $V=[0,\;1]\oplus[0,\;L_c]\times[-1,\;1]$; and the perturbation operator is
\begin{align}\label{per_matrix}
\mathrm{P}= \left[\begin{array}{cccc}  \textrm{P}_{zz} &  \textrm{P}_{zu} &  \textrm{P}_{zp}\\ 
\textrm{P}_{uz} &   \textrm{P}_{uu}& \textrm{P}_{up}\\
\textrm{P}_{pz}&  \textrm{P}_{pu}&   \textrm{P}_{pp} \end{array} \right].
\end{align}
In \citet{Magri2013} we interpreted the perturbation operators $\textrm{P}_{uu}$, $\textrm{P}_{up}$, $\textrm{P}_{pu}$, $\textrm{P}_{pp}$ as possible passive feedback mechanisms (structural sensitivity) and then investigated the base-state sensitivities through $\textrm{P}_{pu}$, $\textrm{P}_{pp}$. 
In this paper we analyse $\mathrm{P}_{pz}$, which is the coupling between the flame and the energy equation (base-state sensitivity), and $\mathrm{P}_{zu}$, which is the coupling between the velocity and the flame equation (structural sensitivity). 
\cccc{$\mathrm{P}_{pz}$ is regarded as a base-state perturbation because it represents a small modification to the flame parameters, such as $Pe$ or $Z_{sto}$ (\S~\ref{rad}). $\mathrm{P}_{zu}$ is regarded as a structural perturbation because it represents a small modification in the intrinsic thermo-acoustic feedback mechanism, in this case between the acoustic velocity and the flame equation (\S~\ref{sec:rec_adv}).}
\par
In this thermo-acoustic system,
there are base-state parameters
for the acoustics and for the flame.
The former, 
which were investigated in \citet{Magri2013},
are the acoustic damping, $\zeta$,
and the flame position, $x_f$. 
\cccc{(The sensitivity of this thermo-acoustic system to perturbations of these parameters is qualitatively identical to that of \citet{Magri2013} because the acoustic models and the direct and adjoint acoustic eigenfunctions are very similar.)}
The latter,
which are new to this study,
are the P\'eclet number, $\Pen$,
the stoichiometric mixture fraction, $Z_{sto}$,
the half width of the fuel slot, $\alpha$,
and the heat-release parameter, $\beta_T=1/T_{av}$, which is the inverse of the average flame temperature. 
\section{Results}
\label{sec_res}
\par
We calculate the
global modes, 
Rayleigh index, 
receptivity, 
and sensitivity 
of two marginally stable/unstable thermo-acoustic systems:
(i) an under-ventilated (open) flame with 
$Z_{sto}=0.125$, $\Pen=35$, $c_1=0.005$, $c_2=0.0065$; 
and 
(ii) an over-ventilated (closed) flame with 
$Z_{sto}=0.8$, $\Pen=60$, $c_1=0.0247$, $c_2=0.018$. 
Both systems have $\alpha=0.35$ and $T_{av}=2/1.316 = 1.520$. 
\cccc{We use $M=225\times N=50$ flame modes and $K=20$ acoustic modes. In appendix \ref{app:convergence} the numerical convergence is shown.
The dominant eigenvalue of the open flame is 
$\sigma=0.00088 + 3.1487\mathrm{i}$ with no temperature jump 
and 
$\sigma=-0.00279 + 5.0938\mathrm{i}$ with a temperature jump of $T_2/T_1=5$.
The closed flame has 
 $\sigma= 0.00408 + 3.1710\mathrm{i}$ with no temperature jump
 and
$\sigma= -0.00756 + 5.1046\mathrm{i}$ with a temperature jump of $T_2/T_1=5$. 
The sets of parameters for $T_2/T_1=1$ have been found to be marginally stable also with the nonlinear code of \citet{Illingworth2013}, which uses a Chebyshev method for the flame and a Galerkin method for the acoustics with no temperature jump. The dominant portion of the spectrum and pseudospectrum of the open-flame case is shown in figure \ref{fig:convergence4} in appendix \ref{app:convergence}.}
%
%
\subsection{The direct eigenfunction (global mode)}
\label{sec_dir_glob_mod}
\begin{figure}
\begin{center}
\includegraphics[width=0.9\textwidth, draft = false]{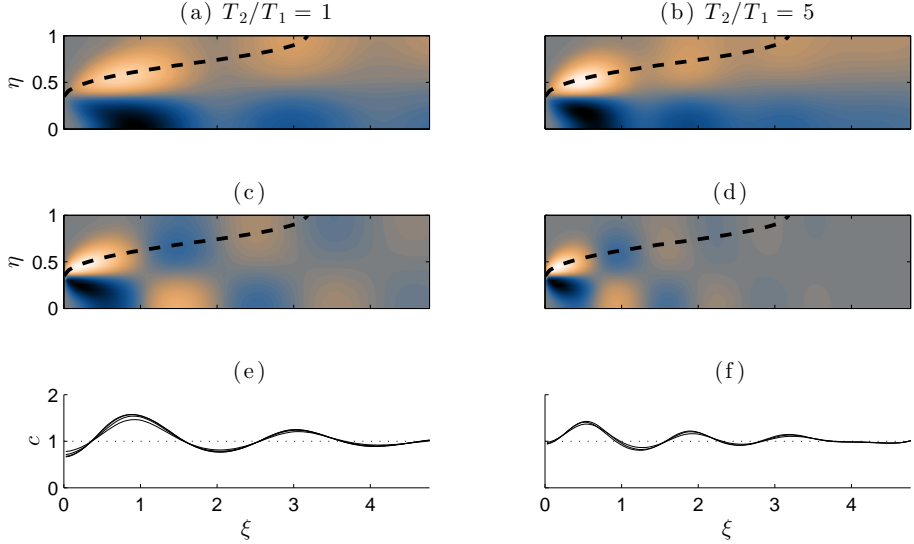}
\caption{(Colour online) Dominant direct eigenfunction (a,b,c,d) and local phase speed (e,f)  of the open flame coupled with acoustics. Results of the left/right column are obtained without/with mean-flow temperature jump. Panels (a,b) show the real parts of the mixture-fraction eigenfunction, panels (c,d) show the imaginary parts. Red/blue colour corresponds to positive/negative value. The dashed line is the steady-flame position. The acoustic component of the eigenfunction is not shown here. Panels (e,f) show the local phase speed, $c$, of the mixture-fraction travelling wave obtained via a Hilbert transform. In (e,f), the solid lines show the phase speed at different cross-stream locations, while the dotted line is the uniform mean-flow speed. The local phase speed is close but not exactly equal to the mean-flow speed.} 
\label{fig:of_z}
\end{center}
\end{figure}
\par
Figures \ref{fig:of_z}a,b,c,d show 
the real and imaginary parts of the direct eigenfunctions of the open flames 
with $T_2/T_1 = 1$ (left) and $T_2/T_1 = 5$ (right). \cccc{The corresponding Galerkin coefficients, $\hat{G}_{n,m}$, are plotted in figure \ref{fig:convergence2} in appendix \ref{app:convergence}.}
The real and imaginary parts are in spatial quadrature,
which shows that the mixture fraction perturbation, $\hat{z}$, 
takes the form of a travelling wave
that moves down the flame in the streamwise direction. 
Panels \ref{fig:of_z}e,f show
the local phase speed of the wave in the streamwise direction,
which is calculated via a Hilbert transform.
(Each solid line corresponds to a different cross-stream location,
showing that the phase speed varies only slightly in the cross-stream direction.) 
In both cases, 
the average phase speed is slightly greater than the mean-flow speed, which is 1. 
This shows that a simple model of the flame,
in which mixture fraction perturbations convect down the flame at the mean-flow speed,
is a reasonable first approximation.
The validity of this approximation increases as the P\'eclet number increases (not shown here)
because convection becomes increasingly more dominant than diffusion.
It is worth noting that the magnitude of $\hat{z}$ decreases in the streamwise direction. This is because the reactants diffuse into each other relatively quickly at this P\'eclet number.
The influence of the mean-flow temperature jump can be seen 
by comparing the direct eigenfunctions 
without (figures~\ref{fig:of_z}a,c) 
and 
with temperature jump (figures~\ref{fig:of_z}b,d). 
When the temperature jump is present, the oscillatory pattern has a shorter wavelength because the frequency of the coupled thermo-acoustic system is higher (figure~\ref{Jump}a). 

\par
In both flames, the mixture fraction perturbation starts at the upstream boundary
and causes heat-release fluctuations when it reaches the flame. 
To the first approximation described above, 
the time delay between the velocity perturbation and the subsequent heat-release perturbation scales with $L_{f}/U$, where $L_{f}$ is the length of the steady flame and $U$ is the mean-flow speed (which is $1$ in this paper).
The phase delay between the velocity perturbation and the subsequent heat release perturbation therefore scales with $L_{f}\sigma_i/U$, where $\sigma_i$ is the dominant eigenvalue's imaginary part, i.e. the linear-oscillation angular frequency. 
We will return to this model and this scaling in the following sections.
%
%
\subsection{The Rayleigh index}\label{ray_viva}
The Rayleigh criterion states that the energy of the acoustic field can grow over one cycle if
$\oint_T \int_{V}p \dot{q} \;\mathrm{d}V \mathrm{d} t \;$
exceeds the damping,
where $V$ is the flow domain and $\mathit{T}$ is the period.
The spatial distribution of 
$\oint_T p \dot{q} \;\mathrm{d} t$, 
which is known as the Rayleigh index, 
reveals the regions of the flow that contribute most to the Rayleigh criterion
and, therefore, gives insight into the physical mechanisms contributing to the oscillation's energy. 
We consider the undamped eigenproblem of the momentum \eqref{equ_gov_mom_ndim} and energy \eqref{equ_gov_enr_ndim} equations. 
Then we multiply the former by $\hat{u}^*$ and the latter by $\hat{p}^*$ and add them up to give
\begin{align}\label{eq:acen_rayn}
2\sigma E_{ac}(x)- \hat{p}^*\hat{q}\delta(x-x_f) = -\left(\hat{u}^*\frac{\partial \hat{p}}{\partial x} + \hat{p}^*\frac{\partial \hat{u}}{\partial x}\right),
\end{align}
where $E_{ac}=1/2\left(\hat{u}^*\hat{u}+\hat{p}^*\hat{p}\right)$ is the thermo-acoustic eigenfunction's acoustic energy. 
Integration of \eqref{eq:acen_rayn} over the flame domain, $[0, L_c]\times[-1, 1]$, and the acoustic domain, $[0, 1]$, gives
\begin{align}\label{eq:acen_rayn2}
2\sigma (2L_c) E_{ac,t}- \int_{-1}^1\int_0^{L_c}\hat{p}_f^*\hat{q}\;\mathrm{d}\xi\mathrm{d}\eta = -(2L_c)\int_0^1\left(\hat{u}^*\frac{\partial \hat{p}}{\partial x} + \hat{p}^*\frac{\partial \hat{u}}{\partial x}\right)\;\mathrm{d}x,
\end{align}
where $E_{ac,t}$ is the (total) acoustic energy, i.e. $E_{ac}$ integrated over the acoustic domain. 
Applying the acoustic boundary conditions, which in this model preclude energy loss at the boundaries (\S \ref{acmodel}),
the real part of \eqref{eq:acen_rayn2} gives
\begin{align}\label{eq:acen_rayn3}
\sigma_r  E_{ac,t} =  \frac{1}{4L_c}\int_{-1}^1\int_0^{L_c}\mathrm{Re}(\hat{p}_f^*\hat{q})\;\mathrm{d}\xi\mathrm{d}\eta. 
\end{align}
By Green's theorem applied to the mixture-fraction equations \eqref{eq:Zst},\eqref{eq:z1}, the right-hand side of \eqref{eq:acen_rayn3} can be expressed as
\begin{align}\label{eq:heat_green}
 \int_{-1}^1\int_0^{L_c}\mathrm{Re}(\hat{p}_f^*\hat{q})\;\mathrm{d}\xi\mathrm{d}\eta&= \nonumber \\
 &\mathrm{Re}\Big\{\hat{p}^*_f\Big[-\frac{1}{Pe}\int_{-1}^{+1}\left(\frac{\partial \hat{z}}{\partial \xi}\right)_{\xi=0}\mathrm{d}\eta\;+ 
\hat{u}_f\int_{-1}^{+1}\hat{z}\left(L_c,\eta\right)\mathrm{d}\eta\;+\nonumber \\
&+\hat{u}_f\bar{Q}\;
-\frac{1}{Pe\left(1-Z_{sto}\right)}\oint_C\nabla \hat{z}\cdot\mathbf{n}\mathrm{d}s\Big]\Big\},
\end{align}
where
$C$ is the curve enclosing the steady stoichiometric line, $\left(\xi_{sto},\eta_{sto}\right)$, and the fuel side; 
$\mathrm{d}s$ is the curvilinear coordinate along the stoichiometric line; 
and the linearized unit-vector normal to the stoichiometric line is:
\begin{align}\label{eq:normal}
\mathbf{n} = 
\left[
\left(
\frac{\partial\bar{Z}}{\partial \xi}
\right)^2 
+ 
\left(
\frac{\partial\bar{Z}}{\partial \eta}
\right)^2
\right]^{-1/2}
\left(\frac{\partial\bar{Z}}{\partial \xi},\frac{\partial\bar{Z}}{\partial \eta}\right).
\end{align}
\par
\begin{figure}
\begin{center}
\includegraphics[width=0.9\textwidth, draft = false]{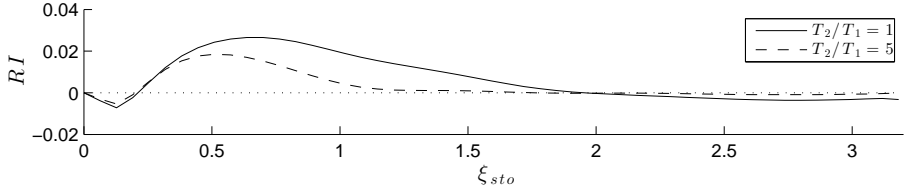}
\caption{The Rayleigh index for the open flame shown as a function of distance along the flame contour $\xi_{sto}$. This shows the part of the flame that most contributes to the increase (positive RI) or decrease (negative RI) in energy of the oscillation over a cycle. The RI reaches a maximum around $\xi_{sto} = 0.5$ to $0.75$ and then decreases because the mixture fraction fluctuations diffuse out as they are convected downstream.}
\label{fig:of_zp00}
\end{center}
\end{figure}

Mixture-fraction fluctuations are created at the base of the flame
by the velocity fluctuations. 
They then convect downstream and cause heat-release fluctuations when they meet the flame.
The influence of these fluctuations depends on their phase relative to the pressure,
as described by the Rayleigh index,
which is the part of \eqref{eq:heat_green} 
that spatially varies in the flame domain 
\begin{align}\label{eq:RI}
RI=\mathrm{Re}\left(-\hat{p}^*_f\nabla \hat{z}\cdot\mathbf{n}\right),
\end{align}
Figure \ref{fig:of_zp00} shows the Rayleigh index
as a function of the distance along the flame, $\xi_{sto}$.
The Rayleigh index reveals that
mixture-fraction perturbations induced by the system itself
(as opposed to those induced by external control in the next section)
most influence the growth rate of thermo-acoustic oscillations 
in the upstream part of the flame $0 < \xi_{sto} < 1.5$.
This is because $\nabla \hat{z}$ is steepest there,
so the rate of species diffusion, and hence reaction rate, is largest there. 
The magnitude and sign of this influence depends on $\xi_{sto}$ 
because the phase relationship between heat release and pressure varies
as the perturbations convect down the flame. 
The Rayleigh index will be compared with maps derived from receptivity and sensitivity analysis in the following sections.
%
%
%
%
\subsection{Receptivity to species injection} \label{sec:rec_inj} 
\begin{figure}
\begin{center}
\includegraphics[width=0.9\textwidth, draft = false]{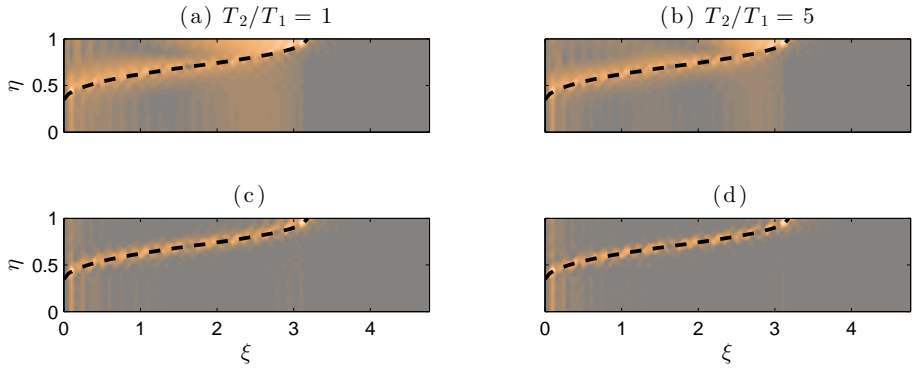}
\caption{(Colour online) Absolute value of the dominant adjoint eigenfunction (a) without and (b) with mean-flow temperature jump. This is for the same operating conditions as those of the direct eigenfunction in figure~\ref{fig:of_z} of the open flame. This is a map of the eigenvalue's receptivity to open-loop forcing via species injection into the mixture-fraction field. It has high amplitude along the flame because species injected into the flame directly affects the reaction rate. It has highest amplitude at the flame tip because the mixture fraction perturbations of the unforced mode have small amplitude at the tip, so the injected species has a proportionately large influence. \cccc{Panels (c,d) show the (dominant) left singular modes, which \ccc{here} correspond to the optimal initial conditions for a final state at $t=0.5$.}} 
\label{fig:of_zp}
\end{center}
\end{figure}
\par
A receptivity analysis 
creates a map in the flame domain
of the first eigenfunction's receptivity to species injection \citep{MagriIJSCD}.
This is given by the adjoint eigenfunction (the adjoint global mode). 
It shows the most effective regions at which 
to place an open-loop active device to excite  the dominant thermo-acoustic mode. 
We imagine perturbing the $z$-field \eqref{eq:z1} on the right-hand side with a forcing term 
that is localized in space:
\begin{equation}\label{rec_adv}
\delta z \; \delta(\xi-\xi_0,\eta-\eta_0)\sin(\omega_st),%
\end{equation}
where 
$\delta z$ is the amount of species injected;
$\delta(\xi-\xi_0,\eta-\eta_0)$ is the Dirac (generalized) function to localize the injection in space at $(\xi_0, \eta_0)$;
and $\omega_s\approx\sigma_i$ is the forcing angular frequency.
\begin{figure}
\begin{center}
\includegraphics[width=0.9\textwidth, draft = false]{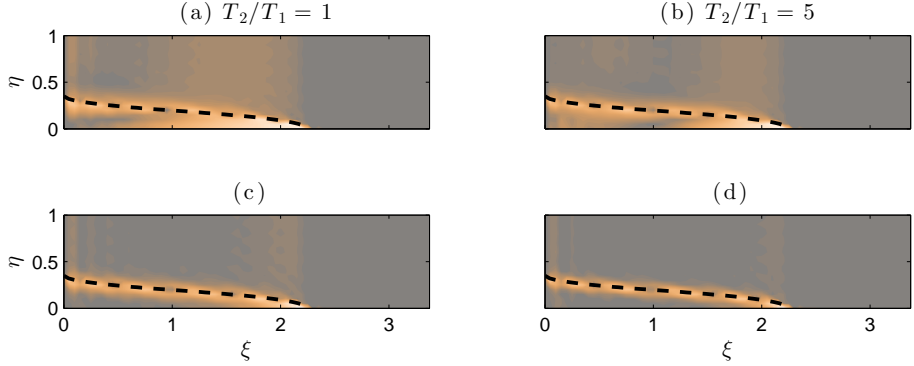}
\caption{As for \ref{fig:of_zp} but for the closed flame. The features are the same as for the open flame.} 
\label{fig:sub_zp}
\end{center}
\end{figure}
\par
The adjoint eigenfunction (figure \ref{fig:of_zp}a,b) 
has high magnitude around the flame. (\cccc{The corresponding Galerkin coefficients, $\hat{G}_{n,m}^+$, are plotted in figure \ref{fig:convergence3} in appendix \ref{app:convergence}.})
This is because species injection affects the heat release only
if it changes the gradient of $\hat{z}$ at the flame itself,
which is achieved by injecting species around the flame.
Its magnitude increases towards the tip of the flame,
where $\nabla \hat{z}$ is weakest.
It is worth comparing this with the Rayleigh index (figure \ref{fig:of_zp00}),
which is greatest towards the base of the flame, where $\nabla \hat{z}$ is strongest.
This reveals that the influence of this particular open-loop control strategy is strongest
at flame positions where the intrinsic instability mechanism is weakest. 
This is because mixture fraction fluctuations diffuse out as they convect down the flame, 
which means that open-loop forcing has a proportionately large influence on the mixture fraction towards the tip.
From a practical point of view, 
this shows that 
open-loop control of the mixture fraction 
has little influence
at the injection plane
but great influence
at the flame tip. 
In this case, this could be achieved by injecting species at the wall. 

\cccc{To check the physical significance of the adjoint eigenfunctions, which in principle live in a different space from those of the direct eigenfunctions, we compare them with the left singular modes, which live in the same space as the direct eigenfunctions.  
On the one hand, the adjoint eigenfunction is the optimal initial condition / forcing maximizing the $L_2$-norm of the thermo-acoustic state in the limit $t\rightarrow\infty$ (see e.g. \citet{MagriIJSCD}).
On the other hand, the left singular mode is the optimal initial condition maximizing the $L_2$-norm of the thermo-acoustic state over a finite time, $t<\infty$. 
Figures \ref{fig:of_zp}c,d show the (dominant) left singular modes for a final state at $t=0.5$. 
Mathematically, these are the (dominant) left singular modes of the propagator $\exp(\mathrm{L}t)$ \citep[see e.g.][]{Schmid2007,Schmid2014}, where L is the linearized thermo-acoustic operator (L=M$^{-1}$B, see\eqref{genop}), evaluated at $t=0.5$.
As expected, the adjoint eigenfunction's shape is very similar to that of the left singular mode.}
There is no substantial difference when the temperature jump is considered (figure~\ref{fig:of_zp}b).
\par
We also investigate the case of a closed flame, 
shown in figure~\ref{fig:sub_zp}. 
Qualitatively, the receptivity is similar to that of the open flame: the system is most sensitive to forcing along the flame and at the flame's tip.
In this case, however, the flame tip lies along the centreline, not along the wall. 
This makes a species-injection strategy more difficult
unless it could be performed, for example,
by injecting droplets that evaporate and burn when they hit the flame's tip
at the centreline. 
%
%
\subsection{Sensitivity to base-state perturbations}\label{rad}
\begin{figure}
\begin{center}
\includegraphics[scale=0.65]{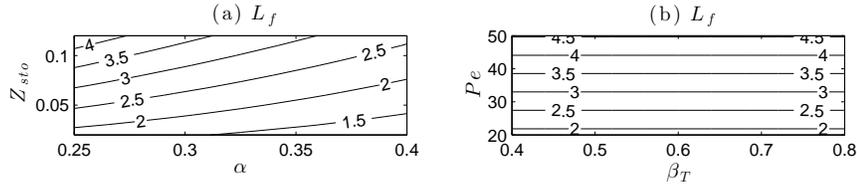}
\caption{Unperturbed steady flame length, $L_{f}$, as a function of (a) the fuel slot half width, $\alpha$, and the stoichiometric mixture fraction, $Z_{sto}$; and (b) the heat-release parameter, $\beta_T$, and the P\'eclet number, $Pe$. $L_f$ is the same for all values of $T_2/T_1$.}
\label{fg:lst}
\end{center}
\end{figure}
\par
The base-state sensitivity analysis
quantifies how the dominant eigenvalue of the thermo-acoustic system, $\sigma$,
is affected by first-order changes to 
$\Pen$, $Z_{sto}$, $\alpha$, and $\beta_T$.
The eigenvalue drift is
\begin{equation}\label{ed_par}
\delta\sigma = \left(\frac{\delta\sigma}{\delta\alpha}\right) \delta\alpha+ 
\left(\frac{\delta\sigma}{\delta\Pen}\right) \delta\Pen+
\left(\frac{\delta\sigma}{\delta Z_{sto}}\right) \delta Z_{sto}+
\left(\frac{\delta\sigma}{\delta\beta_T}\right) \delta\beta_T, 
\end{equation}
in which the terms in brackets are the (complex) sensitivities. 
When $\Pen$, $\alpha$, $Z_{sto}$ are perturbed,
$\bar{Z}$ (\ref{eq:Zst}) changes,
which changes the steady flame shape, which then changes the eigenvalues.
The derivatives of $\bar{Z}$ with respect to $\Pen$, $\alpha$, and $Z_{sto}$
can be evaluated analytically 
because
$\bar{Z}$ has an analytical solution (\ref{zsgb}).
The heat-release parameter of the flame, $\beta_T$, does not directly affect $Z$, 
as can be inferred from \eqref{eq:Z}.
However, it directly affects the amount of heat that feeds into the acoustics (\ref{equ_gov_enr_ndim})
and therefore changes the growth rate without changing the flame shape directly. 
\par
To evaluate the influence of base-state perturbations via \eqref{str_sens_s},
we choose $\Vert \delta \mathrm{C}\Vert\sim O(10^{-6})$, 
which is sufficiently small for nonlinearities to be negligible \citep{Illingworth2013}. 
This was checked by repeating the analysis with a smaller perturbation, $\Vert \delta \mathrm{C}\Vert\sim O(10^{-7})$, for which \cccc{the real and imaginary parts of} the eigenvalues changed by $\sim O(10^{-9})$.
We analyse the sensitivities around marginally stable points: 
$\delta\sigma/\delta Z_{sto}$, $\delta\sigma/\delta\alpha$ in the range \cccc{$Z_{sto}=[0.02, 0.12]$} and $\alpha=[0.25, 0.4]$; 
and
$\delta\sigma/\delta Pe$ and $\delta\sigma/\delta\beta_T$ in the range $Pe=[20, 50]$ and $\beta_T=[0.4, 0.8]$.
The sensitivities are calculated with $T_2/T_1=1$ and $T_2/T_1=5$. 
In the following analysis,
the length of the unperturbed flame emerges as a key parameter.  
This is defined here as the distance between the inlet and the tip of the steady flame.
It is shown as a function of $Z_{sto}$ and $\alpha$ in figure \ref{fg:lst}a 
and as a function of $Pe$ and $\beta_T$ in figure \ref{fg:lst}b.
The flame length increases 
as $Z_{sto}$ increases, as $\alpha$ decreases, and as $Pe$ increases,
but is not a function of $\beta_T$ or $T_2/T_1$.
%
%
\begin{figure}
\begin{center}
\includegraphics[scale=0.65]{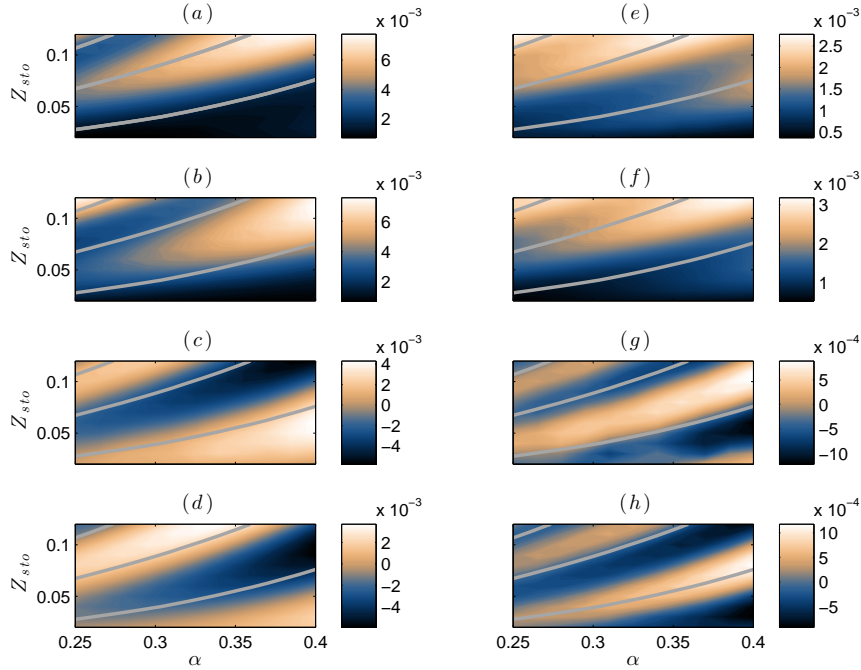}
\caption{(Colour online) Sensitivities to base-state perturbations of $\alpha$ and $Z_{sto}$. $T_2/T_1=1$ left column, $T_2/T_1=5$ right column with the steady-flame length contours of figure \ref{fg:lst} superimposed. The sensitivities depend strongly on $Z_{sto}$ and $\alpha$ but are similar at similar values of $L_f$.
}
\label{fg:BaseStateOF}
\end{center}
\end{figure}
\begin{figure}
\begin{center}
\includegraphics[scale=0.65]{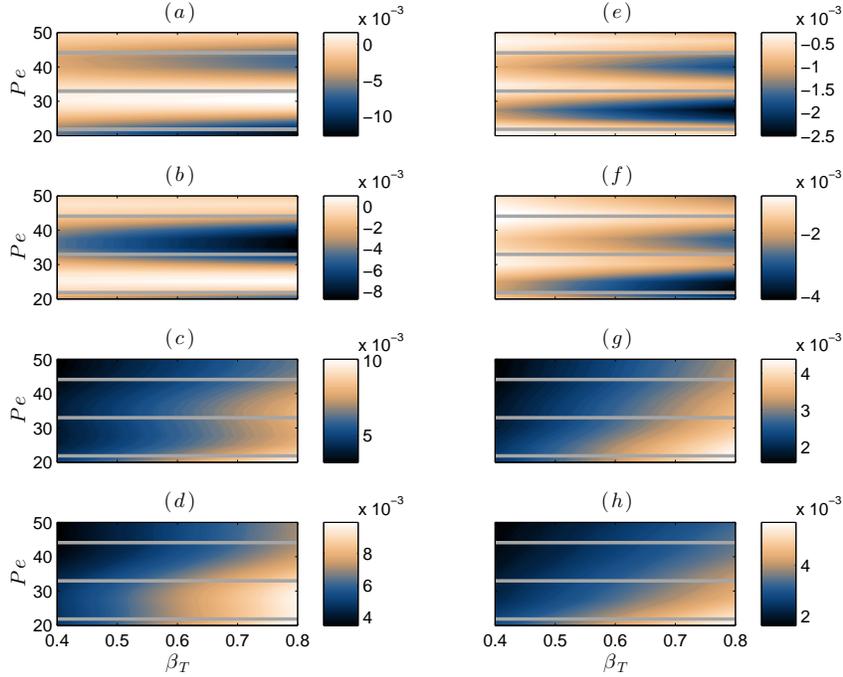}
\caption{(Colour online) Sensitivities to base-state perturbations of $Pe$ and $\beta_T$. $T_2/T_1=1$ left column, $T_2/T_1=5$ right column with the steady-flame length contours superimposed. The sensitivities $\delta\sigma/\delta Pe$ depend strongly on $Pe$ but not $\beta_T$.}
\label{fg:BaseStateOFII}
\end{center}
\end{figure}
\begin{figure}
\begin{center}
\includegraphics[scale=0.65]{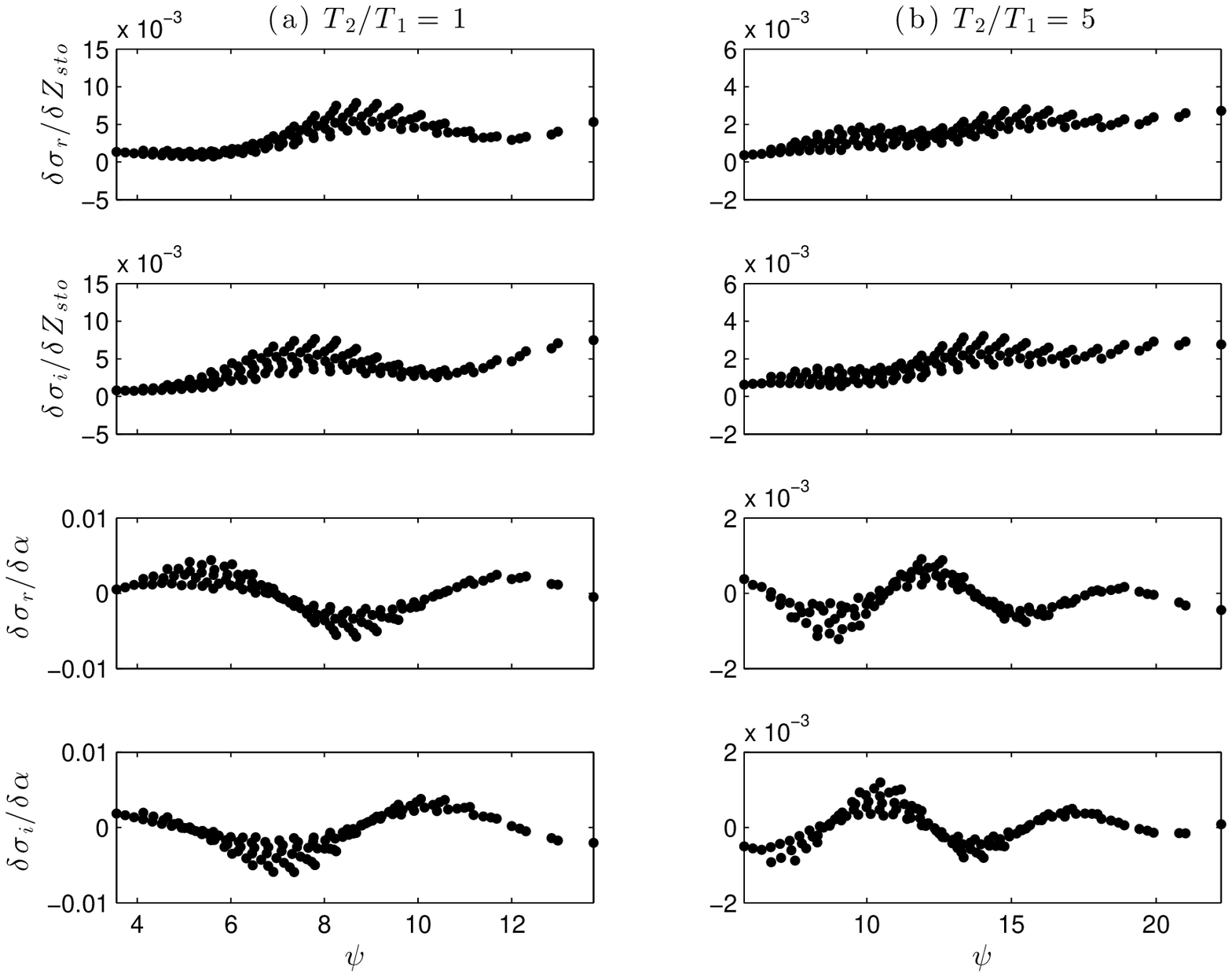}
\caption{The data from figure \ref{fg:BaseStateOF} plotted as a function of the phase between pressure and heat release oscillations, as estimated by $\psi \equiv L_{f}\sigma_i / U$. Solutions (a) with no temperature jump, (b) with temperature jump.
\textcolor{black}{The data does not collapse exactly to a curve because
perturbations in $z$ do not convect down the flame exactly at speed $U$.}}
\label{fig:BaseStateColl}
\end{center}
\end{figure}
\par
The change of the growth rate, $\sigma_r$, and the frequency, $\sigma_i$, 
due to small changes in $Z_{sto}$ and $\alpha$ are shown in figure \ref{fg:BaseStateOF}
and those due to small changes in $Pe$ and $\beta_T$ in figure \ref{fg:BaseStateOFII}. 
Changes in $Z_{sto}$ can be achieved by diluting the fuel or oxidizer. 
As shown by \eqref{Zsto}, 
$Z_{sto}$ increases when
the oxidizer mass fraction, $X_i$, increases
or the fuel mass fraction, $Y_i$, decreases.
Changes in $Pe$ are achieved by adjusting the mean-flow velocity
(see appendix \ref{appscale}), as long as the mean-flow Mach number is small.
These results, obtained by an adjoint-based approach, have been checked against the solutions obtained via finite difference and agree to within $\sim O(10^{-9})$.

\par
These figures are useful from a design point of view. 
For example, they reveal that 
at $Z_{sto} = 0.12$ and $\alpha = 0.38$,
changes in $Z_{sto}$ strongly influence the growth rate
but that 
at $Z_{sto} = 0.11$ and $\alpha = 0.40$,
changes in $Z_{sto}$ strongly influence the frequency instead.
This demonstrates an inconvenient feature of thermo-acoustic instability:
the influence of each parameter is exceedingly sensitive to small changes in the base state (i.e. the operating point). 

\par
\textcolor{black}{
It can be seen that
$\delta \sigma / \delta Z_{sto}$,
$\delta \sigma / \delta \alpha$, and
$\delta \sigma / \delta Pe$,
oscillate in spatial quadrature in parameter space
(e.g. 
local maxima of  $\delta \sigma_r / \delta Z_{sto}$
lie between local maxima of 
$\delta \sigma_i / \delta Z_{sto}$
and vice-versa).
Furthermore, 
lines of constant
$\delta \sigma / \delta Z_{sto}$,
$\delta \sigma / \delta \alpha$, and 
$\delta \sigma / \delta Pe$
very nearly follow the lines of constant $L_{f}$ shown in figure \ref{fig:BaseStateColl}.
}
\par
\textcolor{black}{
These observations can be explained physically by considering the simple criterion of the thermo-acoustic instability mechanism described in \S\ref{sec_dir_glob_mod}.
In this criterion, the velocity perturbations cause $z$ perturbations at the base of the flame.  These are convected downstream and cause a heat-release perturbation some time later. 
This time delay, $\tau$, scales with $L_{f} / U$, where $L_{f}$ is the length of the flame. 
The influence of this heat-release perturbation 
depends on the phase of the heat release relative to the phase of the pressure (for the growth rate) or velocity (for the frequency),
which are in temporal quadrature. 
This is why the base-state sensitivity plots are in spatial quadrature in parameter space.
The oscillatory pattern is not observed for $\delta \sigma / \delta \beta_T$
because $\beta_T$ affects only the heat release at the flame and not the steady flame length
and therefore has no direct influence on the phase delay.
}
\par
\textcolor{black}{
The phase delay, $\psi$, is given by $\tau/T$, where $T = 2\pi/\sigma_i$.
In this simple model, $\delta \sigma$ depends only on $\psi$,
which means that, if the simple model were sufficient, 
the eigenvalue drifts in figures 
\ref{fg:BaseStateOF} and \ref{fg:BaseStateOFII} 
would collapse onto a single curve
when plotted as a function of $\psi=L_{f} \sigma_i / U$.
This is shown in figure \ref{fig:BaseStateColl} 
for $\delta \sigma_r / \delta Z_{sto}$ and $\delta \sigma_i / \delta Z_{sto}$ as a function of $\psi$ for (a) $T_2/T_1 = 1$ and (b) $T_2/T_1 = 5$. 
The data at each $T_2/T_1$ collapse somewhat closely to a curve,
particularly for $\delta \sigma / \delta \alpha$. 
The data does not collapse exactly because
perturbations in $z$ do not convect down the flame at a uniform speed,
as shown in figures \ref{fig:of_z}e,f,
and the flame length, $L_f$, 
is a simplistic measure of the change in shape of the flame 
caused by changes in $Z_{sto}$, $\alpha$, and $Pe$.
Nevertheless, this simple criterion is useful for physical understanding,
while the data in figures \ref{fg:BaseStateOF} and \ref{fg:BaseStateOFII}
shows the influence of base-state modifications exactly.
}
%
%
%
\subsection{Structural sensitivity to species advection fluctuations} \label{sec:rec_adv} 
By inspection of the governing equation of the perturbed $z$ field (\ref{eq:z1}), 
we can interpret the term $-u_f\partial\bar{Z}/\partial\xi$ as 
an intrinsic forcing of $z$ due to advection in the streamwise direction. 
In this section 
we perform a structural sensitivity analysis using the framework in \S\ref{sec:str_th}  
in order to reveal the locations where a small change in the advective velocity field
most influences the eigenvalue of the thermo-acoustic system through this term. 
This can be loosely interpreted as the location of the core of the thermo-acoustic instability, which can then be compared with the Rayleigh index.

\cccc{The structural perturbation to the flame-velocity field
is assumed to be localized in the flame domain}:
\begin{equation}\label{rec_adv}
\delta \textrm{P}= -\delta C_{zu}\hat{u}_f\frac{\partial \bar{Z}}{\partial \xi}\delta(\xi-\xi_0,\eta-\eta_0),
\end{equation}
where $\delta C_{zu}$ is the small perturbation coefficient,  
and $\delta(\xi-\xi_0,\eta-\eta_0)$ is the Dirac (generalized) function, 
which reproduces the impulsive effect of the perturbation at $(\xi_0, \eta_0)$.
\cccc{Note that such a flame-velocity perturbation occurs at the acoustic flame location, $x=x_f$, because the flame is a pointwise source for the acoustics (see figure \ref{model}). Therefore, the structural perturbation \eqref{rec_adv} is naturally localized in the acoustic domain.}
Following \eqref{per_matrix}, 
the perturbation operator representing feedback proportional to the acoustic velocity and entering the flame equation is $\textrm{P}_{zu}=-\hat{u}_f\partial \bar{Z}/\partial \xi$.
The overlap of $\hat{z}^{+*}$ and $-\hat{u}_f\partial \bar{Z}/\partial \xi$ gives a map of the flame's sensitivity to small changes in the velocity field:
\begin{align}\label{sens_str_u}
\frac{\delta\sigma}{\delta C_{zu}}= \frac{-\hat{z}^{+*}\hat{u}_f \frac{\partial \bar{Z}}{\partial \xi}}{ \int_{V}
\left[\hat{z}^{+*};\hat{u}^{+*};\hat{p}^{+*}\right] \cdot
\left[\hat{z};\hat{u};\hat{p} \right]
 \;\mathrm{d} V }.
\end{align}
\begin{figure}
\begin{center}
\includegraphics[width=0.9\textwidth, draft = false]{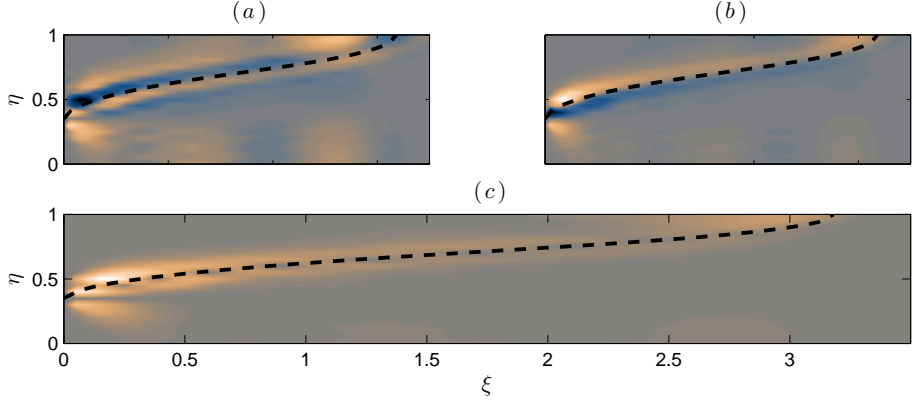}
\caption{(Colour online) Real (a), imaginary (b) and absolute (c) values of the structural sensitivity of the open flame with $T_2/T_1=5$. This shows where the eigenvalue of the thermo-acoustic system is most sensitive to changes in the advective velocity field.}
\label{fig:OF_SS_xieta}
\end{center}
\end{figure}
\begin{figure}
\begin{center}
\includegraphics[width=0.9\textwidth, draft = false]{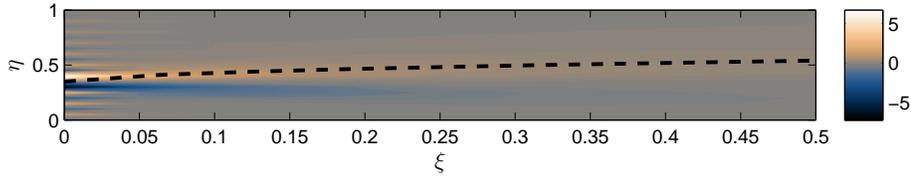}
\caption{(Colour online) Axial gradient of the steady mixture fraction, $\partial\bar{Z}/\partial\xi$. It has high amplitude near the inlet plane.} 
\label{fig:GradStoXi}
\end{center}
\end{figure}
This is shown in figure \ref{fig:OF_SS_xieta}. 
It is worth noting that 
the adjoint eigenfunction, $\hat{z}^{+}$ (figure \ref{fig:of_zp}a,b) 
has highest amplitude near the flame tip, 
that $\hat{u}_f$ is uniform,
and that $\frac{\partial \bar{Z}}{\partial \xi}$ 
has highest amplitude near the flame base (figure \ref{fig:GradStoXi}),
where the steady mixture-fraction axial gradient is greatest.
These combine to give the structural sensitivity, ${\delta\sigma}/{\delta C_{zu}}$.
This shows that changes to the velocity field have most influence 
(i) at the flame, 
because changes in velocity advection there directly change the reaction rate,
as did the open-loop species injection in \S \ref{sec:rec_inj};
(ii) in the region $0 < \xi < 1$
which, as expected, is the region in which the Rayleigh index is large (figure \ref{fig:of_zp00}).

The structural sensitivity also shows where a passive feedback device 
would have most influence on the eigenvalue. 
For example, 
the drag from a small cylinder 
generates a negative perturbation to the velocity, $\delta C_{zu}<0$.
The first-order effect of such a cylinder has no influence on the steady flame (the base flow) because its equation \eqref{eq:Zst} is linear. This means that the presence of a small cylinder changes the eigenvalue of the thermo-acoustic system only through the unsteady $z$ field. (This structural sensitivity analysis is simple, 
because the momentum equation is not solved in the flame domain.)
When placed in the blue region of figure \ref{fig:OF_SS_xieta}a,
this perturbation would destabilize the thermo-acoustic system 
and when placed in the red region it would stabilize it. 

\section{Conclusions}
\label{sec_conc}
%
\par 
The main goal of this paper is 
to apply adjoint sensitivity analysis to a low-order thermo-acoustic system.
Our first application of this analysis \citep{Magri2013}
was to an electrically heated Rijke tube
with an imposed time-delay between velocity fluctuations and heat-release fluctuations.
Our application in this paper
is to a diffusion flame in a duct.
The model and its discretization originate from \citet{Balasubramanian2008,Magri}, 
which was recently revised \citep{Magri2013d}.
The model contains 
a diffusion flame with infinite-rate chemistry
coupled with one-dimensional acoustics in an open-ended duct.
It includes the effect of the mean-flow temperature jump at the flame. 
Rather than impose a time-delay between velocity and heat release fluctuations,
we model convection and reaction in the flame domain.
This provides a more accurate representation of the thermo-acoustic system
and the base-state variables that influence its stability,
which are the main focus of this paper. 
\par
We use adjoint equations to calculate 
the system's 
receptivity to species injection, 
sensitivity to base-state perturbations,
and structural sensitivity to advective-velocity perturbations.
We compare these with the Rayleigh index.
We derive the continuous adjoint equations for completeness 
but we use the discrete adjoint approach for the calculations
because it is easier and more accurate for this application. 

The receptivity to species injection
reveals that the thermo-acoustic system is most receptive
to open-loop forcing of the mixture fraction
towards the tip of the flame.
This is because
mixture-fraction fluctuations 
diffuse out as they convect down the flame. 
Consequently, 
open-loop forcing has a proportionately large influence on the mixture fraction towards the tip of the flame.
For the same reason, the Rayleigh index is small there. 
The receptivity map is useful 
when designing open-loop strategies for control/excitation 
of thermo-acoustic oscillations.
Without performing a receptivity analysis,
it may not be obvious that the flame is most sensitive to forcing of the mixture fraction at positions along the flame where the Rayleigh index is small. 

\par
The sensitivity to base-state perturbations
reveals the sensitivity to perturbations in the combustion parameters, 
which in this case are 
the stoichiometric mixture fraction, $\delta Z_{sto}$;
the fuel slot to duct width ratio, $\alpha$; 
the P\'eclet number, $\Pen$; 
and the heat-release parameter, $\beta_T$. 
Although these can be found with classical finite difference calculations, 
using the adjoint equations significantly reduces the number of computations
without affecting the accuracy. 
Overall, the thermo-acoustic system is most sensitive to changes in 
$\delta Z_{sto}$, $\delta\beta_T$ and $\delta\alpha$, 
but least sensitive to $\delta\Pen$.
As expected,
these sensitivities depend strongly on the phase delay between 
velocity perturbations and subsequent heat release perturbations.
This phase delay scales with $L_f \sigma_i / U$,
where $L_f$ is the flame length, 
$U$ is the flow speed,
and $\sigma_i$ is the oscillation angular frequency.
The stoichiometric mixture fraction, $Z_{sto}$,
and fuel slot with, $\alpha$,
change the flame length.
These are the easiest parameters to change in an experiment,
although control with these would be delicate 
because of the sensitivity's oscillatory patterns (figure \ref{fg:BaseStateOF}).
The inverse of the average flame temperature, $\beta_T$, 
changes the influence of the flame's heat release.
If this can be changed, then control with this is attractive
because $\beta_T$ does not directly affect the flame length 
and therefore the sensitivity to this parameter 
does not oscillate (figure \ref{fg:BaseStateOFII}).
The P\'eclet number, $\Pen$ 
has very little influence for most of the operating points considered in this paper. 
Even if it could be changed, it would not be a useful parameter for passive control. 
The base state sensitivity analysis also reveals a feature that seems to be common to all thermo-acoustic systems: the influence of base state parameters is exceedingly sensitive to small changes in the operating point. 

\par
The structural sensitivity 
shows the effect that a generic advection-feedback mechanism would have
on the frequency and growth rate of the thermo-acoustic oscillations. 
It can be loosely interpreted as the location of the core of the thermo-acoustic instability. 
This structural sensitivity analysis is simple, 
because the momentum equation is not solved in the flame domain.
Nevertheless, it shows
(i) the regions in which a passive control device is most effective at controlling the thermo-acoustic oscillations;
(ii) the regions where future velocity models must capture the species advection most accurately. 
As expected, the structural sensitivity is large in regions in which the Rayleigh index is large. 

\par
This paper shows that 
adjoint receptivity and sensitivity analysis 
can be applied to 
thermo-acoustic systems that simulate the flame,
as well as to 
those that impose a time delay between velocity and heat-release fluctuations
\citep{Magri2013}.  
With very few calculations, 
this analysis reveals how
each parameter affects the stability of a thermo-acoustic system,
which is useful information for practitioners. 
Although many technical challenges remain,
this analysis can be extended to more accurate models,
particularly those that simulate the velocity field in the flame domain,
and is a promising new tool for the analysis and control of thermo-acoustic oscillations. 

\par
The authors would like to thank  Prof. R. I. Sujith and Dr. K. Balasubramanian for providing their code used in \citet{Kulkarni2011} and their comments on the Galerkin method applied to the flame. The authors are grateful to Dr. S. Illingworth, Dr. I. C. Waugh and Dr. O. Tammisola for helpful discussions. 
This work is supported by the European Research Council through Project ALORS 2590620.

\appendix
\section{Scale factors for non-dimensionalization}\label{appscale}
Dimensional quantities are denoted with  $\tilde{}$. 
The acoustic variables are scaled as: 
$\tilde{L}_{a}x =\tilde{x}$ [m], 
$\tilde{L}_at_{ac}/\tilde{c}_1=\tilde{t}$ [s], 
$\tilde{u}_1u=\tilde{u}$ [m/s], 
$\tilde{\rho}_1 \rho = \tilde{\rho}$ [kg/m$^3$], 
$\gamma M_1 \tilde{p}_1p=\tilde{p}$ [Pa]; 
where 
$\tilde{L}_a$ [m] is the length of the duct, 
$\tilde{c}_1$ [m/s] is the speed of sound in the cold mean-flow, 
$\tilde{u}_1$ [m/s] is the cold mean-flow velocity, 
$\tilde{\rho}_1$ [kg/m$^3$] is the cold mean-flow density, 
$\tilde{p}_1$ [Pa] is the mean-flow pressure, 
$\gamma=\tilde{c}_p/\tilde{c}_v$, and 
$M_1$ is the cold mean-flow Mach number. 
$\tilde{c}_p$ and $\tilde{c}_v$ are the mass heat capacities at constant pressure and constant volume of the mixture [J kg$^{-1}$K$^{-1}$]. %

The combustion variables are scaled as: 
$\tilde{H}\xi =\tilde{\xi}$ [m], 
$\tilde{H}\eta =\tilde{\eta}$ [m], 
$\tilde{H}t_c/\tilde{u}_1=\tilde{t}$ [s], 
$\tilde{T}_{ref}T=\tilde{T}$ [K],
where $\tilde{T}_{ref} = (Y^*_i\tilde{Q}_h)/\tilde{c}_p$, and
$\tilde{Q}_h$ is the heat released by combustion of 1 kg of fuel [J kg$^{-1}$ /Y$_i$$^*$] \citep{Poinsot2005}. 
The combustion time scale has been chosen 
to be exactly the same as
the acoustic time scale, i.e. $t_{ac}=t_c$. 
This can be achieved provided that $M\tilde{L}_c/\tilde{H}=1$ (compact flame and low Mach number assumptions). The non-dimensional length of the combustion domain along $\xi$ is $L_c = \tilde{L}_c/\tilde{H}$. The P\'eclet number is the ratio between the diffusion and convective time scales, $\Pen = \tilde{u}_1\tilde{H}/\tilde{\mathcal{D}}$, where $\tilde{\mathcal{D}}$ is the (uniform) mass-diffusion coefficient [m$^{2}$ s]. 
\section{Effect of the mean flow on the acoustic frequencies}\label{appMeanFlow}
\begin{figure}
\begin{center}
\includegraphics[width=0.7\textwidth, draft = false]{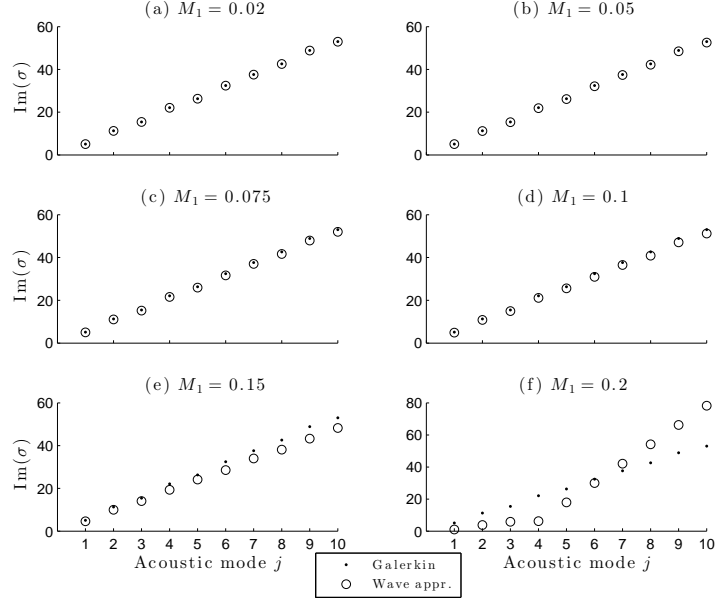}
\caption{Comparison between the acoustic angular frequencies, Im($\sigma$), calculated by the wave approach (circles) and the Galerkin method (dots). $M_1$ is the Mach number of the mean flow upstream of the flame.} 
\label{fig:compWG}
\end{center}
\end{figure}

The acoustic angular frequencies obtained by the wave approach (see online supplementary material for details) are shown in figure \ref{fig:compWG} and compared with the angular frequencies calculated via the Galerkin method \eqref{eig_val}. 
The effect of the mean-flow velocity, which is neglected in the Galerkin formulation, becomes influential for mean-flow Mach numbers $\gtrsim 0.1$.
\section{Steady flame solution}\label{ssbs}
\begin{figure}
\begin{center}
\includegraphics[width=0.7\textwidth, draft = false]{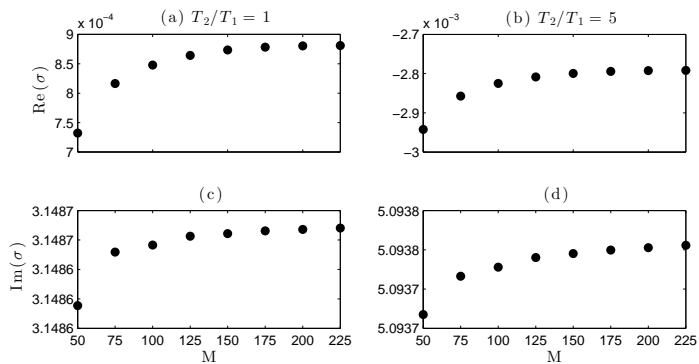}
\caption{Dominant eigenvalue convergence with respect to the number of axial Galerkin flame modes, M. Transversal Galerkin flame modes and acoustic modes are fixed to $N=50$ and $K=20$, respectively. The system's parameters are those of the open-flame case (see \S \ref{sec_res}).} 
\label{fig:convergence}
\end{center}
\end{figure}

\begin{figure}
\begin{center}
\includegraphics[width=0.7\textwidth, draft = false]{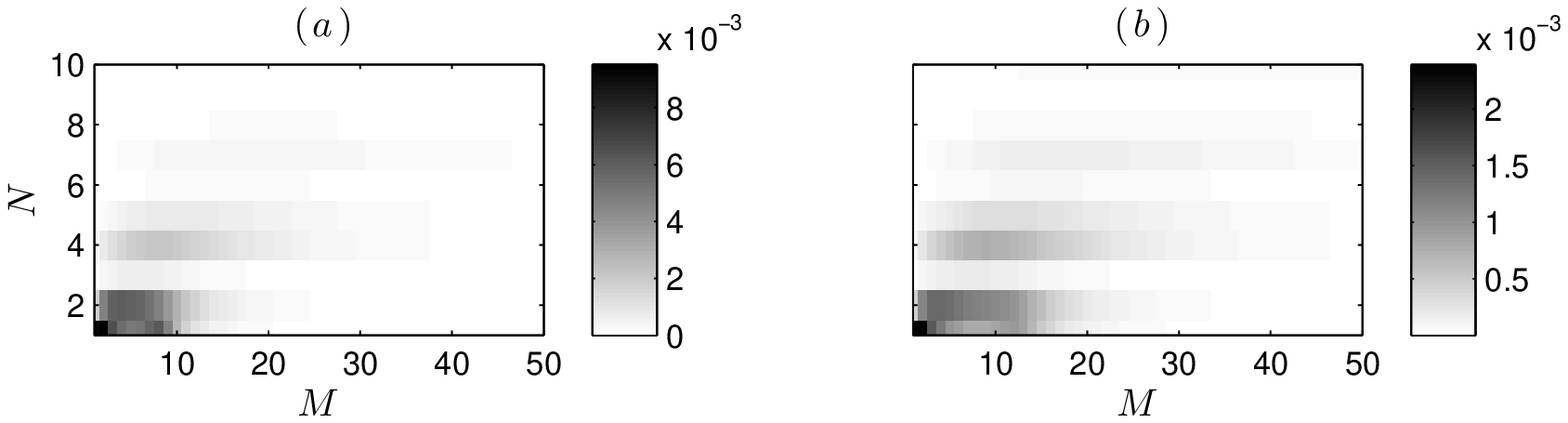}
\caption{Absolute value of the direct Galerkin coefficients $\hat{G}_{n,m}$, of the open flame.} 
\label{fig:convergence2}
\end{center}
\end{figure}

\begin{figure}
\begin{center}
\includegraphics[width=0.7\textwidth, draft = false]{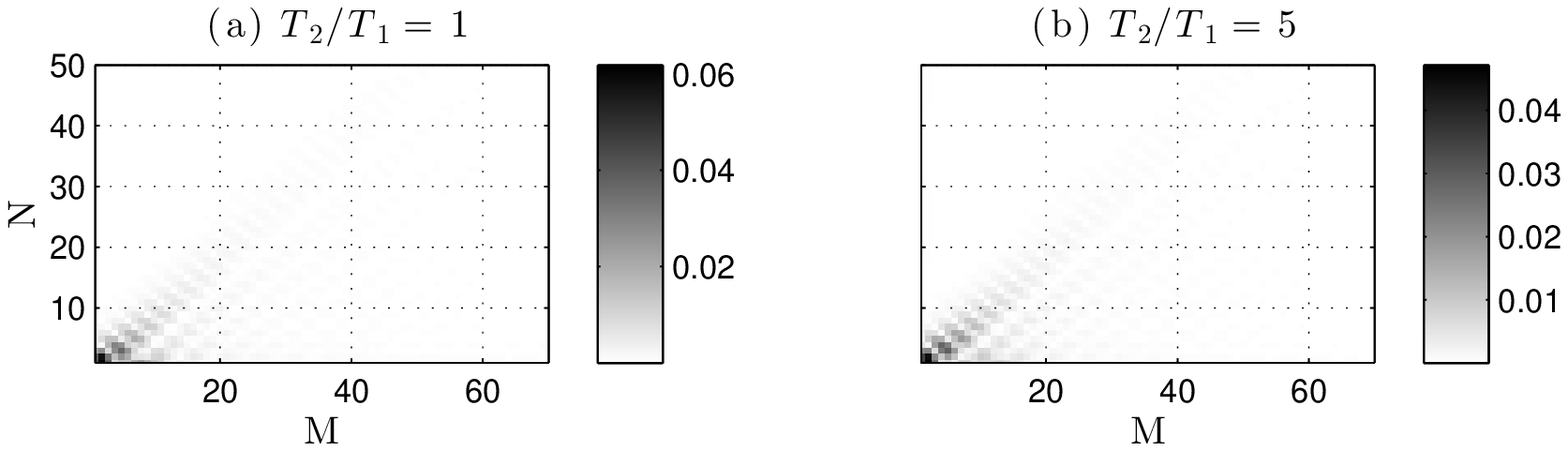}
\caption{Absolute value of the adjoint Galerkin coefficients, $\hat{G}_{n,m}^+$, of the open flame.} 
\label{fig:convergence3}
\end{center}
\end{figure}

\begin{figure}
\begin{center}
\includegraphics[width=0.7\textwidth, draft = false]{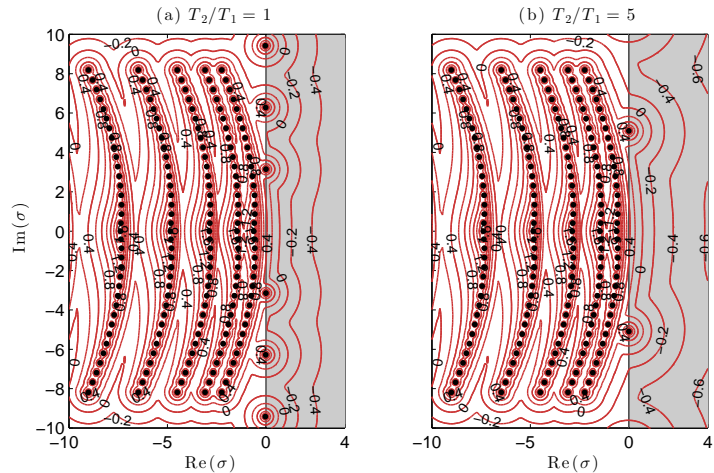}
\caption{Dominant portion of the spectrum and pseudospectrum of the open flame.} 
\label{fig:convergence4}
\end{center}
\end{figure}

The steady flame solution is obtained by separation of variables \citep{Magri2013d,Magri2013e}:
\begin{equation}\label{zsgb}
\bar{Z}=\alpha
+\frac{2}{\upi}\sum_{n=1}^{+\infty}\frac{\sin(n\upi\alpha)}{n\left(1+b_n\right)}\cos(n\upi \eta)\left[\exp(a_{n1} \xi)+b_n \exp(a_{n2} \xi)\right],
\end{equation}
where 
\begin{align}
&a_{n1}\equiv\frac{\pe}{2}-\sqrt{\frac{\pe^2}{4}+n^2\upi^2}, \;\;\;a_{n2}\equiv\frac{\pe}{2}+\sqrt{\frac{\pe^2}{4}+n^2\upi^2},\\
&b_n\equiv-\frac{a_{n1}}{a_{n2}}\exp\left(-2 L_c \sqrt{\frac{\pe^2}{4}+n^2\upi^2}\right).
\end{align}
Note that if $L_c\rightarrow\infty$, then $b_n\rightarrow0$. 
In this limit, (\ref{zsgb}) coincides with the solution proposed by \citet[eq.~(7), p. 966]{Magina2013}. 
(Note that they defined the characteristic convective scale for the P\'eclet number, $Pe$, to be $\alpha \tilde{H}$.)
 \cccc{ 
 \section{Numerical convergence, spectrum and pseudospectrum}\label{app:convergence}
We use as many Galerkin modes as required to obtain numerical convergence of the direct/adjoint dominant eigenvalues and eigenfunctions. 
A numerical discretization of $M=225$ $\times$ $N=50$ flame modes and $K=20$ acoustic modes achieves such a convergence. Figure \ref{fig:convergence} shows the convergence rate of the dominant eigenvalue for the open-flame system without temperature jump (left panels) and with
temperature jump (right panels). The relative errors are Re($\sigma_{M=225}-\sigma_{M=200})$/Re($\sigma_{M=200})\sim O(10^{-4})$ and Im($\sigma_{M=225}-\sigma_{M=200})$/Im($\sigma_{M=200})\sim O(10^{-7})$. 
When $M=225$, $N=75$ and $K=30$ modes are used, the relative errors are  $\sim O(10^{-4})$, for the growth rate, and $\sim O(10^{-10})$, for the angular frequency. We therefore used $M=225$, $N=50$ and $K=20$ as a good compromise between accuracy and computational time. Similar accuracy has been obtained for the closed-flame case. 
The most significant Galerkin coefficients of the direct and adjoint eigenproblems, \eqref{genop} and \eqref{genopa}, are depicted in figures \ref{fig:convergence2} and \ref{fig:convergence3}, respectively. These figures show that the most energetic direct and adjoint modes are concentrated in the first modes, and the Galerkin coefficients decrease as the mode indices increase.  Finally, the dominant portion of the spectrum and pseudospectrum of the open-flame case is shown in figure \ref{fig:convergence4}. The pseudospectra are nearly concentric circles centred on the eigenvalues even when the temperature jump in modelled (figure \ref{fig:convergence4}b). This means that this thermo-acoustic system is weakly non-normal, in agreement with \citet{Magri2013d}, regardless of the temperature jump. }

\bibliographystyle{jfm}
\bibliography{library}
\end{document}